\documentclass[iop,apj]{emulateapj}      

\usepackage{amsmath,amssymb,amsxtra,amsfonts}
\usepackage{graphicx}
\usepackage{epsfig}
\usepackage{natbib}
\usepackage{txfonts}

\newcommand{\da}{D_{\rm A}}
\newcommand{\dac}{D_{\rm A}^{\rm cluster}}
\newcommand{\dl}{D_{\rm L}}

\shorttitle{Morphology of Galaxy Clusters: A Test of the DD Relation}
\shortauthors{Meng et al.}

\begin{document}

\title{Morphology of Galaxy Clusters: A Cosmological Model-Independent Test of the Cosmic Distance-Duality Relation}

\author{Xiao-Lei Meng\altaffilmark{1,3}, Tong-Jie Zhang\altaffilmark{1,2}, Hu Zhan\altaffilmark{3} and Xin Wang\altaffilmark{4,5}}
\email{tjzhang@bnu.edu.cn}

\altaffiltext{1}{Department of Astronomy, Beijing Normal University, Beijing 100875,
China} \altaffiltext{2}{Center for High Energy Physics, Peking University, Beijing
100871, China} \altaffiltext{3}{Key Laboratory for Optical Astronomy, National
Astronomical Observatories, Chinese Academy of Sciences, Beijing 100012, China}
\altaffiltext{4}{Department of Astronomy, Nanjing University, Nanjing, 210093, China}
\altaffiltext{5}{Key laboratory of Modern Astronomy and Astrophysics (Nanjing
University), Ministry of Education, Nanjing 210093, China}

\begin{abstract}
Aiming at comparing different morphological models of galaxy clusters, we use two new
methods to make a cosmological model-independent test of the distance-duality (DD)
relation. The luminosity distances come from Union2 compilation of Supernovae Type Ia.
The angular diameter distances are given by two cluster models (De Filippis et al. and
Bonamente et al.). The advantage of our methods is that it can reduce statistical errors.
Concerning the morphological hypotheses for cluster models, it is mainly focused on the
comparison between elliptical $\beta$-model and spherical $\beta$-model. The spherical
$\beta$-model is divided into two groups in terms of different reduction methods of
angular diameter distances, i.e. conservative spherical $\beta$-model and corrected
spherical $\beta$-model. Our results show that the DD relation is consistent with the
elliptical $\beta$-model at $1\sigma$ confidence level (CL) for both methods, whereas for
almost all spherical $\beta$-model parameterizations, the DD relation can only be
accommodated at $3\sigma$ CL, particularly for the conservative spherical $\beta$-model.
In order to minimize systematic uncertainties, we also apply the test to the overlap
sample, i.e. the same set of clusters modeled by both De Filippis et al. and Bonamente et
al.. It is found that the DD relation is compatible with the elliptically modeled overlap
sample at $1\sigma$ CL, however for most of the parameterizations, the DD relation can
not be accommodated even at $3\sigma$ CL for any of the two spherical $\beta$-models.
Therefore it is reasonable that the marked triaxial ellipsoidal model is a better
geometrical hypothesis describing the structure of the galaxy cluster compared with the
spherical $\beta$-model if the DD relation is valid in cosmological observations.
\end{abstract}

\keywords{Galaxies: clusters: general --- distance scale
--- X-rays: galaxies: clusters --- cosmic background radiation ---
Cosmology: observations --- supernovae: general}

\section{Introduction}\label{sect:intro}

Galaxy clusters are crucial to our understanding of the universe. For example, galaxy
clusters have been used to derive the Hubble constant
\citep{2000ApJ...533...38R,mason01,cunha07,2010ARA&A..48..673F}, to discriminate between
different cosmological models
\citep{1995ApJ...447....8M,1998MNRAS.296.1061T,suwa03,2006ApJ...647....8H}, to test the
intracluster gas mass distribution and temperature profile
\citep{coo98,2003A&A...398...41P,2007A&A...474..745P,2011MNRAS.tmp..803P}, and to trace
out the thermodynamical history using scaling relations among cluster observables
\citep{2007MNRAS.379..518M,2009MNRAS.400.1085S}. These studies often require modeling of
cluster morphology and hence can benefit from a better understanding of cluster
morphology.

Hypotheses about an object's three-dimensional properties are not easily tested through
two-dimensional observations \citep{2004ApJ...601..599L}. One of the major questions
about cluster morphology is whether it is spherical or triaxial
\citep{2002ApJ...574...38F}. Simulations have shown that dark matter halos are triaxial
\citep{jing02,2005ApJ...629..781K}, and there is evidence from strong
gravitational-lensing observations \citep{2005A&A...443..793G} as well. Moreover, some
efforts have been made to reconstruct three-dimensional cluster morphology and correct
the projection effect using observational data
\citep{2004ApJ...601..599L,2006A&A...455..791P,2007MNRAS.380.1207S}. In this paper, we
examine the spherical and elliptical models of cluster morphology using the cosmic
distance-duality (DD) relation \citep[also called Etherington relation]{ether33}.

The DD relation plays a fundamental role in observational cosmology, covering the
analyses of galaxy cluster observations \citep{cunha07}, gravitational-lensing phenomena
\citep{sch99} and the cosmic microwave background (CMB) data \citep{koma11}. The DD
relation connects different metric distances via
\begin{equation}\label{eq:da-dl}
\dl=\da(1+z)^2
\end{equation}
where $\dl$ and $\da$ are the luminosity distance and angular diameter distance
respectively, with $z$ being the redshift. The DD relation is a general duality in all
metric theories of gravity, as long as there is no sink or source along the null
geodesics.

One way to test the validity of the DD relation is to combine the metric distance results
from both observations and theoretical expressions in a given cosmological model.
\citet{uzan04} proposed the idea of testing the DD relation using $\da$ from X-ray
surface brightness and Sunyaev-Zel'dovich effect \citep[SZE,][]{sze72} measurements of
galaxy clusters. They concluded that there was no significant violation of the DD
relation for a $\Lambda$-Cold Dark Matter ($\Lambda$CDM) model. \citet{deb06} examined
the DD relation with $\da$ of 38 clusters from the \citet{bona06} sample. They defined
$\eta(z)=\dl(z)/\left[\da(z)(1+z)^2\right]$ and found $\eta=0.97\pm0.03$ at 1$\sigma$
confidence level (CL). Thus, there is no violation for the DD relation in the framework
of the $\Lambda$CDM model. With $\dl$ from data compilation of Type Ia supernovae (SNe
Ia), the DD relation was also used to constrain cosmic opacity by \citet{av10}. They
introduced a parameter $\varepsilon$ to analyze the deviation of the DD relation in a
flat $\Lambda$CDM model, by assuming that it satisfies $\dl=\da(1+z)^{2+\varepsilon}$,
and got the constraint, $\varepsilon=-0.04^{+0.08}_{-0.07}$ ($2\sigma$ CL).
\citet{holan11} offered a method for testing the DD relation using \textit{WMAP} (7-year)
results by fixing the $\Lambda$CDM model. Particularly, they considered two different
geometries for galaxy clusters, i.e. elliptical and spherical models. $\da$ of these two
models were derived by the joint analysis of X-ray surface brightness observations plus
SZE data. Their analysis was based on two parametric representations of $\eta(z)$, i.e.
$\eta(z)=1+\eta_0z$ and $\eta(z)=1+\eta_0z/(1+z)$. They concluded that the best-fit value
for the elliptical model is close to $\eta_0=0$ at $1\sigma$ CL, whereas for the
spherical model, the result is only marginally compatible at $3\sigma$ CL.

More robust, another possibility of testing the DD relation is via the combination of
different sets of observations that furnish both metric distances, i.e. $\dl$ and $\da$,
independently. \citet{holan10} made such kind of cosmological model-independent test of
the DD relation with $\da$ from galaxy clusters observations and $\dl$ from Constitution
SNe Ia data. They employed the $\da$ data from two cluster models. The first model was
defined by 25 $\da$ of clusters \citep{defi05} using an elliptical $\beta$-model over the
redshift interval $0.023\le{z}\le0.784$. The second model contained 38 $\da$ of clusters
\citep{bona06} in the redshift range $0.14<z<0.89$ observed by \textit{Chandra} and
\textit{Owens Valley Radio Observatory/Berkeley-Illinois-Maryland Association}
interferometric arrays. For each cluster, they selected one SN Ia which has the closest
redshift to the cluster's, requiring that the difference in redshift ($\Delta z=| z_{\rm
cluster} - z_{\rm SN}|$) is smaller than 0.005. So a direct test for the DD relation
could be allowed. They also referred to $\eta(z)$ as the aforementioned two
representations, and obtained $\eta_0=-0.28^{+0.44}_{-0.44}$ ($2\sigma$ CL) for the
elliptical model and $\eta_0=-0.42^{+0.34}_{-0.34}$ ($3\sigma$ CL) for the spherical
model.

In a subsequent paper, \citet{li11} also performed a cosmological model-independent test
of the DD relation, using the most recent compilation which consists of 557 SNe Ia
\citep[Union2 compilation,][]{amanu10}. Compared with the Constitution set used by
\citet{holan10}, the values of $\Delta z$ are more centered around the $\Delta z=0$ line.
Furthermore, they assumed two more general parameterizations for the test of the DD
relation, i.e. $\eta(z)=\eta_0+\eta_1z$ and $\eta(z)=\eta_0+\eta_1z/(1+z)$. Finally, they
obtained the conclusion that the DD relation can be accommodated at $1\sigma$ and
$3\sigma$ CLs for the elliptical and spherical models with $\eta(z)=1+\eta_0z$ and
$\eta(z)=1+\eta_0z/(1+z)$, whereas $1\sigma$ and $2\sigma$ CLs by postulating two more
general parameterizations for the two models.

There are two aims of this paper: (1) making a cosmological model-independent test of the
DD relation with two new methods using SNe Ia data from Union2 compilation and galaxy
clusters data from two morphological models reported by \citet{defi05} and \cite{bona06},
(2) testing the intrinsic shape of clusters if the DD relation is compatible with present
observations. In this paper, two new methods to obtain $\dl$ at cluster redshifts are
employed: (I) fitting the Union2 SNe Ia data with weighted least-squares and
interpolating $\dl$ at each cluster's redshift, (II) binning SNe Ia $\dl$ within the
redshift range $|z_{\rm cluster} - z_{\rm SN}|<0.005$ to get $\dl$ at the cluster's
redshift. In doing so, we also take into account the asymmetric uncertainties in
\citet{bona06}'s $\da$ data. Furthermore, two cluster morphological models, i.e.
elliptical $\beta$-model and spherical $\beta$-model, are tested. The latter model is
divided into two groups in terms of the different reduction methods of the angular
diameter distance data, i.e. conservative spherical $\beta$-model and corrected spherical
$\beta$-model. To avoid limitations of any particular parameterizations of the deviation
from the DD relation, four one-dimensional and four two-dimensional parameterizations are
applied to the maximum likelihood estimation test.

This paper is organized as follows. In Section 2 we briefly describe the SNe Ia data and
galaxy cluster models. The analysis methods and results are presented in Section 3.
Section 4 gives our conclusions and discussions.

\section{SNe Ia Data and Galaxy Cluster Models}\label{sect:data}

\begin{figure}\label{figure1}
\epsscale{1.25} \plotone{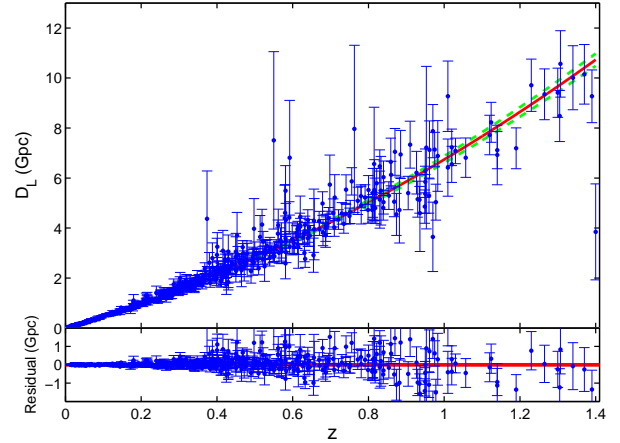} \caption{Quadratic Fit to the Union2 SNe Ia data. Our
best fit and the 1$\sigma$ CL are represented by red solid and green dashed lines
respectively. The blue points with error bars in the upper and lower panel stand for the
Union2 data and their residuals compared with the fitting function.}
\end{figure}

In our analysis, the validity of the DD relation is tested using $\dl$ and $\da$ results
from mutually and cosmologically independent measurements. Two sets of $\da$ data from
X-ray and SZE observations of galaxy clusters are considered. In order to get $\dl$, we
choose the Union2 sample of \citet{amanu10}, which contains 557 well-measured SNe Ia.
Compared with the ``Union'' SNe Ia compilation, the new compilation increases the number
of well-measured distant SNe Ia by including the events discovered in ground-based
searches during 2001 and then followed with the Wide Field Planetary Camera 2 on the
\textit{Hubble Space Telescope}.

An important issue is that there are only eight SNe Ia host galaxies for which
high-quality Cepheid distance measurements are possible \citep{2011ApJ...730..119R}. In
other words, although the sample in Union2 compilation is sufficiently large, the
calibration of SNe Ia still suffers from small number statistical uncertainty, due to
lack of SNe Ia calibrators \citep{2010ARA&A..48..673F}. In order to consider this
uncertain absolute magnitude, we add an uncertainty of 0.05 magnitudes in quadrature as a
covariance among all distance moduli of 557 SNe Ia, provided by \textit{Supernova
Cosmology Project}\footnote{http://supernova.lbl.gov/Union/} \citep{amanu10}.

Both \citet{holan10} and \citet{li11} employed a moderate redshift criterion, $\Delta z=|
z_{\rm cluster} - z_{\rm SN}| <0.005$, and selected the nearest SN Ia for every galaxy
cluster to test the DD relation statistically. However using merely one luminosity
distance $\dl$ from all those available within the same redshift range will lead to
larger statistical errors.\footnote{For instance, there is a galaxy cluster Abell 1413,
at $z=0.142$ \citep{bona06}. Five SNe Ia in the Union2 SNe Ia compilation satisfy the
criterion $|\Delta z|<0.005$: SN2005fj ($z=0.143$), SN2005fw ($z=0.143$), SN1999bm
($z=0.1441$), SN2005ld ($z=0.145$) and SN2005gx ($z=0.146$) \citep{amanu10}. Then how to
choose one from the 2 SNe Ia at $z=0.143$? Using both will improve the statistics. Hence,
as a generalization, we propose to use all SNe Ia $\dl$ within $|\Delta z|<0.005$.}
Instead of using $\dl$ of Union2 SNe Ia directly, two new methods are adopted,

\hspace{0.5cm}(I) fitting the Union2 SNe Ia data and interpolating $\dl$ at each
cluster's redshift;

\hspace{0.5cm}(II) Binning the Union2 SNe Ia $\dl$ within the redshift range $|z_{\rm
cluster} - z_{\rm SN}|<0.005$ to get $\dl$ at the cluster's redshift.

In realizing method (I), a quadratic fit is applied to all the data with errors under the
principle of weighted least-squares fitting (in other words, $\chi^2$ fitting
\citep{press92}). The fitting curve and its $1\sigma$ CL are plotted along with the
original data in Figure~1.

As suggested by others \citep{holan10,li11}, the $|\Delta z|<0.005$ criterion is employed
to select SNe Ia $\dl$ data. But in method (II), we take an inverse variance weighted
average of all the selected data in the following manner. Assuming that $D_{\textrm{L}i}$
represents $i$th appropriate SNe Ia luminosity distance data (within $|\Delta z|<0.005$)
with $\sigma_{D_{\textrm{L}i}}$ denoting its reported observational uncertainty, in light
of conventional data reduction techniques by \citet[Chap.~4]{Bev03}, it is
straightforward to obtain
\begin{equation}
\begin{array}{l}
\bar{\dl}=\frac{\sum\left(D_{\textrm{L}i}/\sigma^2_{D_{\textrm{L}i}}\right)}{\sum1/\sigma^2_{D_{\textrm{L}i}}} ,\\
\sigma^2_{\bar{\dl}}=\frac{1}{\sum1/\sigma^2_{D_{\textrm{L}i}}},
\end{array}\label{eq:dlsigdl}
\end{equation}
where $\bar{\dl}$ stands for the weighted mean luminosity distance at the corresponding
galaxy cluster redshift, and $\sigma_{\bar{\dl}}$ is its uncertainty.

To get reasonable results for cluster $\da$, one has to assume certain cluster
morphologies. The structure of clusters is an essential cosmological probe, as it plays
important roles in discriminating different cosmological models via the mass density of
the universe \citep{rich92,suwa03}, and constraining the halo evolution models
\citep{jing02}. Moreover, different assumptions of cluster shape affect the measurements
of baryon fraction significantly \citep{coo98,allen04}. Generally speaking, a robust way
to measure cluster $\da$ is through the joint analysis of X-ray emission and SZE. The hot
intracluster medium interacts with CMB photons via inverse Compton scattering, causing a
change in the apparent brightness of CMB and a small distortion of CMB spectrum, i.e. the
SZE \citep{sze72,birk99}\citep[also see][for a recent review]{2002ARA&A..40..643C}. This
effect is insensitive to the redshift of galaxy clusters, particularly at high redshifts
($z>1$) \citep{2002ARA&A..40..643C}. In calculating thermal SZE, one should consider the
relativistic corrections. \citet{1998ApJ...502....7I} proposed a convenient analytical
expression up to fifth order terms in $k_{\rm B}T_{\rm e}/m_{\rm e}c^2$. If the peculiar
velocity of clusters is non-negligible, namely the kinematic SZE, the relativistic
corrections are presented by \citet{1998ApJ...508...17N}. Mostly recently,
\citet{2006NCimB.121..487N} improve these corrections to fourth order in $k_{\rm B}T_{\rm
e}/m_{\rm e}c^2$.

Two cluster models under different morphological hypotheses are utilized. The first model
involves 25 X-ray-selected clusters, with measured SZE temperature decrements
\citep{defi05}. A marked triaxial ellipsoidal $\beta$-model is reconstructed to describe
the cluster structure, as described by
\begin{equation}
n_{\rm e}(r)=n_{\rm e0}\left(1+\frac{e_{1}^2 x_{\rm 1,int}^2+e_{2}^2
x_{\rm 2,int}^2 + e_{3}^2 x_{\rm 3,int}^2}{r_{\rm
c}^2}\right)^{-3\beta/2}
\end{equation}
with ${x_{i,\rm int}}~~(i=1,2,3)$ defining the intrinsic orthogonal coordinates aligned
with the three corresponding principal axes. ${e_{i}}~~(i=1,2,3)$ represent the axial
ratios (note that one of them is unity), and $r_{\rm c}$ denotes the core radius along
one of the principal axis. As a combination of \textit{Chandra, XMM-Newton} and
\textit{ROSAT} observations, this data set consists of two sub-samples, one of which is
comprised of 18 galaxy clusters with $0.14\le z\le0.78$ \citep{reese02}. The other
sub-sample, analyzed by \citet{mason01}, contains seven clusters from the X-ray
flux-limited sample of \citet{ebel96}. The second morphological model includes 38 galaxy
clusters in the redshift range from 0.14 to 0.89, provided by \citet{bona06}. They
modeled the electron density profile by a generalized spherical $\beta$-model,
\begin{equation}
n_{\rm e}(r)=n_{\rm e0}\left[f\left(1+\frac{r^2}{r_{\rm
c1}^2}\right)^{-3\beta/2}+(1-f)\left(1+\frac{r^2}{r_{\rm
c2}^2}\right)^{-3\beta/2}\right],
\end{equation}
where $r_{\rm c1}$ and $r_{\rm c2}$ correspond to the two core radii which describe the
inner and outer portions of density distribution respectively, and $f$ is a factor
between zero and unity. The angular diameter distance data of every individual galaxy
cluster with the two models are summarized in Table 1.

As reported by \citet{bona06}, almost all angular diameter distances for the spherical
$\beta$-model are followed by asymmetric uncertainties. According to \citet{d'a04}, the
sources of asymmetric uncertainties include non-Gaussianity of the likelihood curve,
nonlinear propagation, and some systematic effects. And thus, the real value of physical
quantities of interest is biased. \citet{holan10} and \citet{li11} addressed this issue
by combining the statistical and systematic uncertainties in quadrature. As stressed by
\citet{bona06}, the modeling uncertainties of the angular diameter distances presented in
their Table 2, 4 and 5, contribute to statistical uncertainties. Therefore, it is
reasonable to make the following corrections and estimations \citep[equations~(15) and
(16)]{d'a04}, i.e. $\textrm{E}(\da)\approx\da+\textit{O}(\Delta_+-\Delta_-),
~~\sigma_{\da}\approx(\Delta_++\Delta_-)/2$. We also use the reported $\da$ value as the
expected value ($\textrm{E}(\da)$) and the larger flank of each two-sided error
(max($\Delta_+,~\Delta_-$)) as the standard deviation ($\sigma_{\da}$). Hereafter we name
the original spherical $\da$ set the conservative spherical $\beta$-model, and the $\da$
set, which has been applied a moderate compensation shift to, the corrected spherical
$\beta$-model. Results of the angular diameter distances with conservative and corrected
spherical $\beta$-models are also listed in Table 1. Separately, we extract an overlap
sample of clusters that are analyzed by both elliptical and spherical models. The result
from the overlap sample provides a direct comparison of two morphological models of
galaxy clusters.

\section{Analysis Method and Results}

\begin{figure*}\label{figure2}
\epsscale{1.15} \plottwo{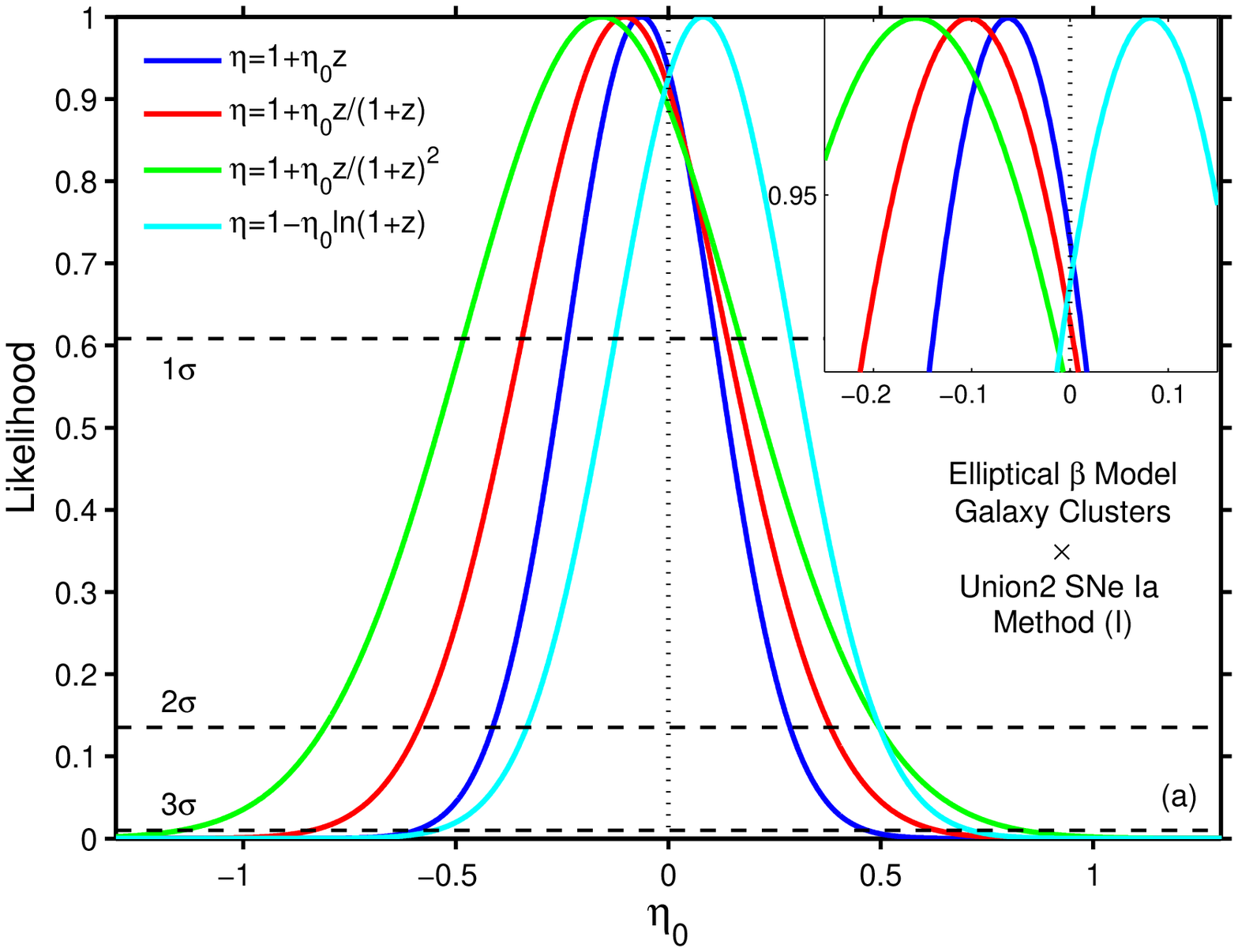}{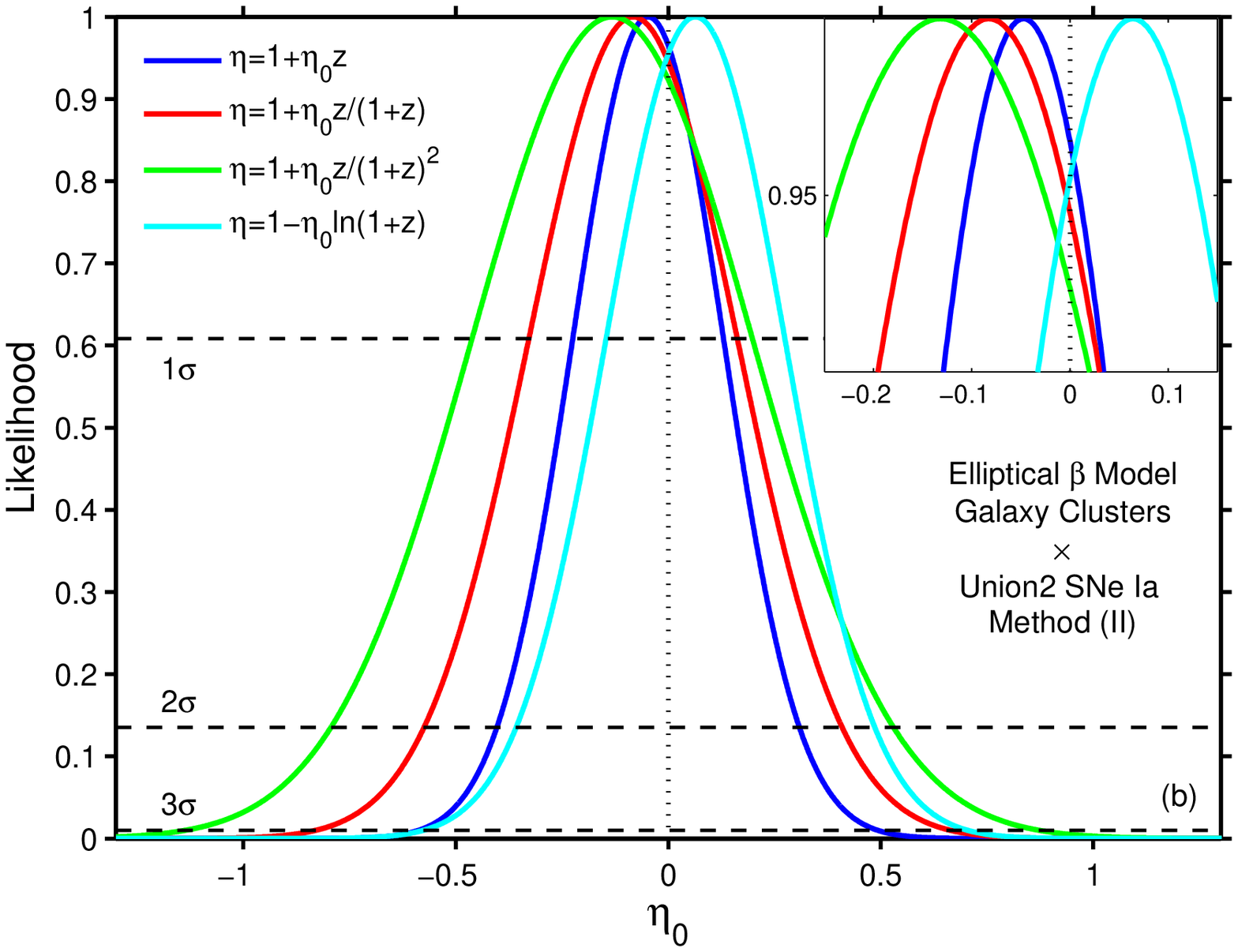} \caption{Likelihood distribution functions
for elliptical $\beta$-model in four one-dimensional parameterizations. Panel (a)
corresponds to the results of SNe Ia data obtained with method (I), and panel (b)
corresponds to those with method (II). The inserts of panel (a) and (b) give zoom-in
views near the peaks of the likelihood functions.}
\end{figure*}

\begin{figure*}\label{figure3}
\epsscale{1.15} \plottwo{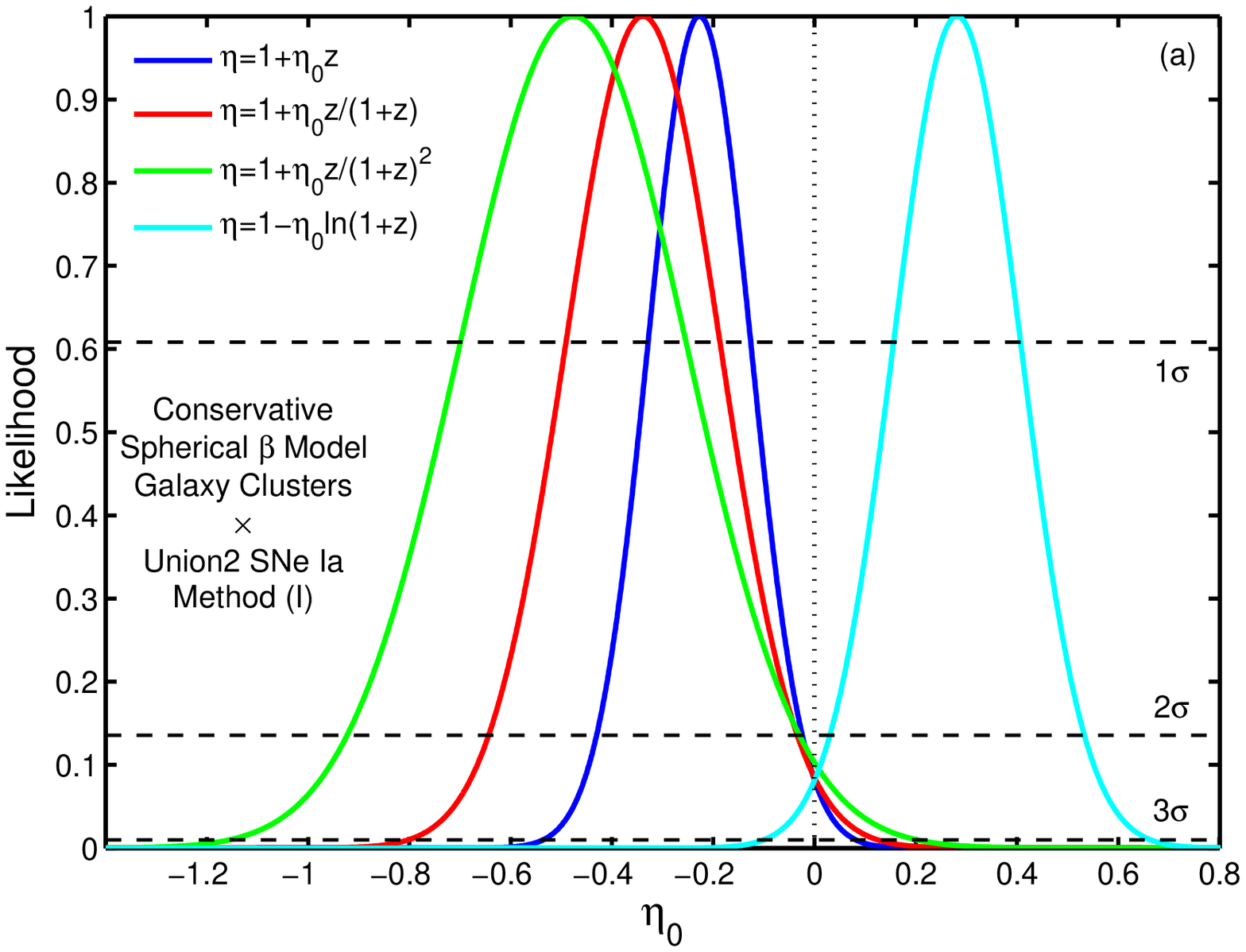}{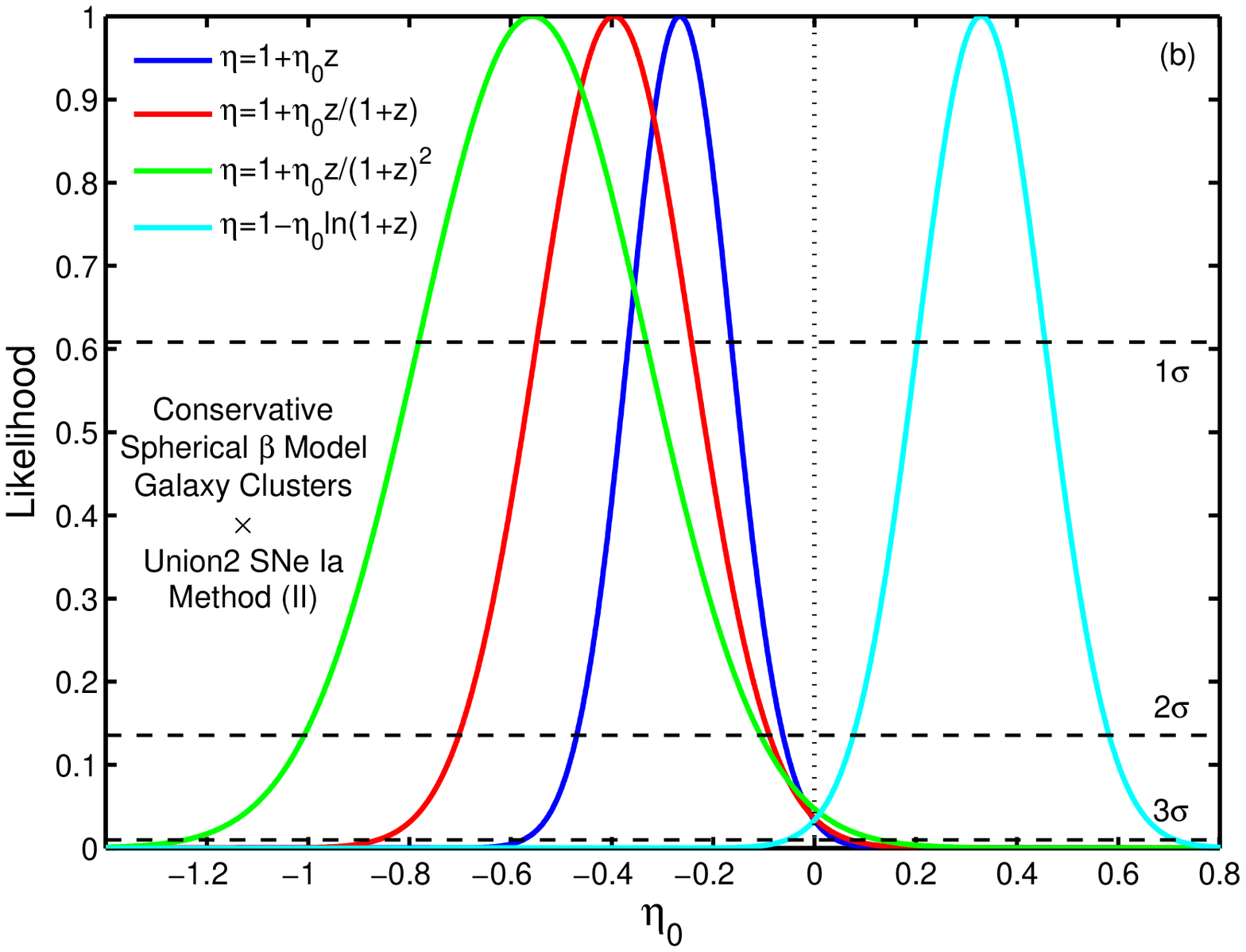} \epsscale{1.15}
\plottwo{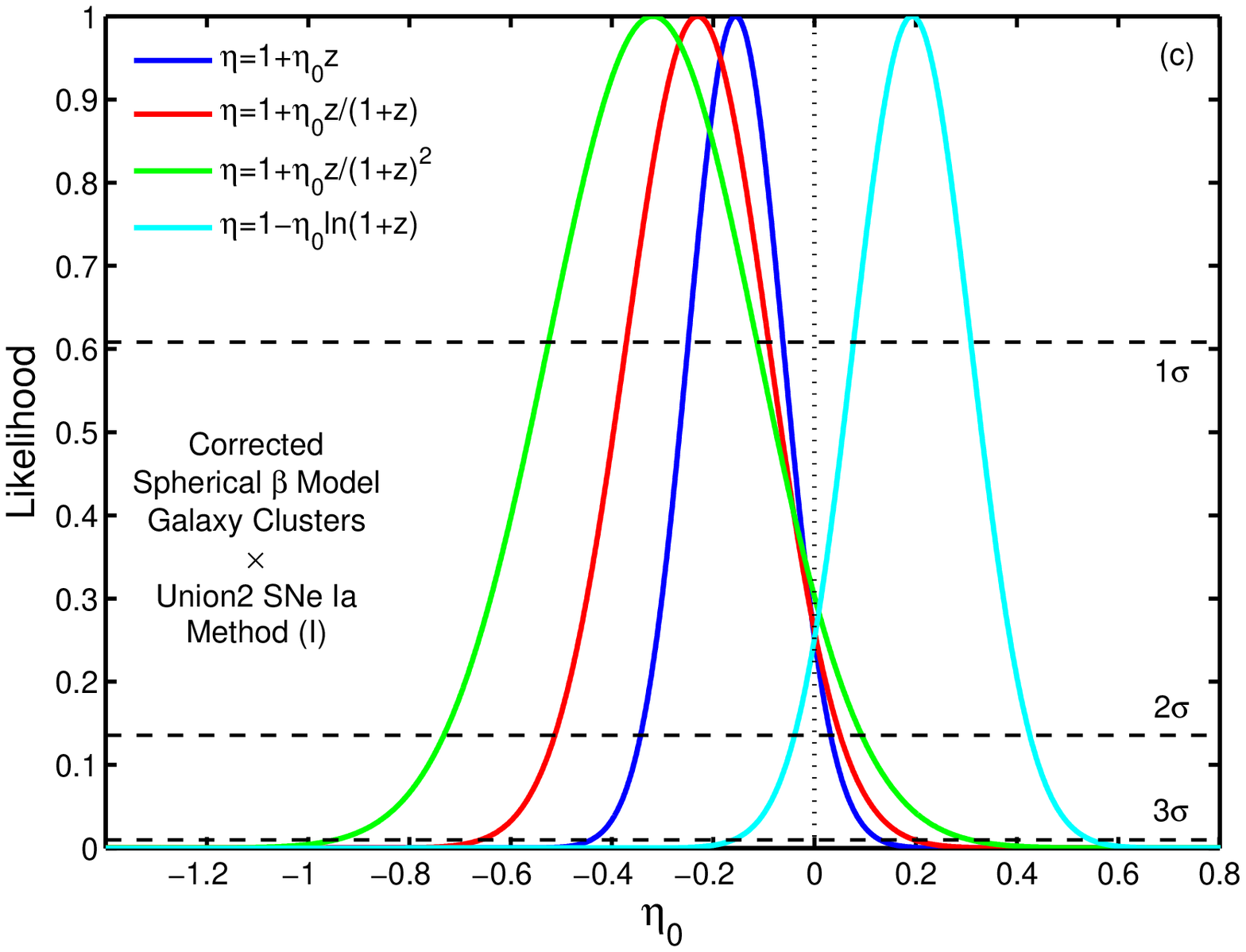}{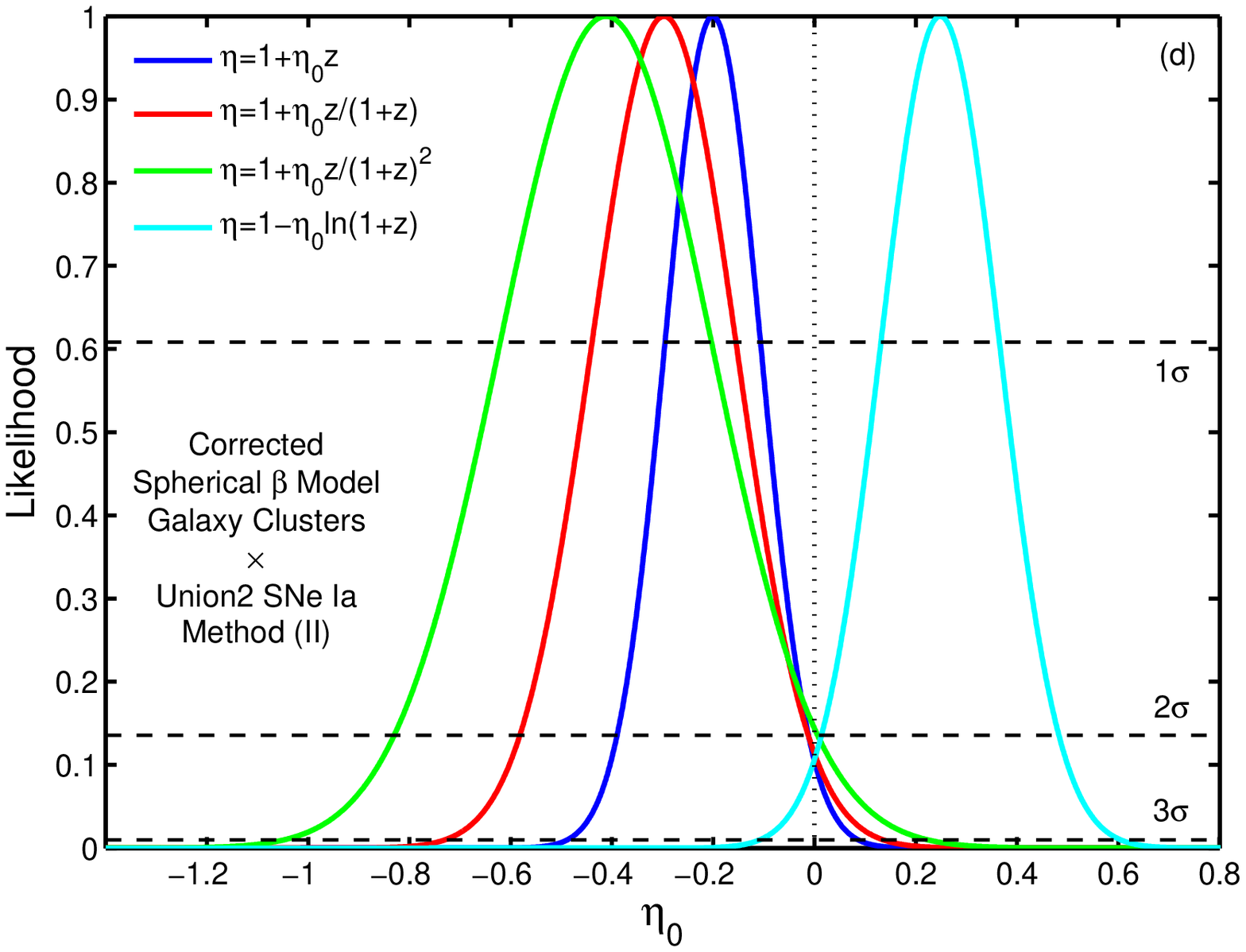} \caption{Likelihood distribution functions for
conservative spherical $\beta$-model and corrected spherical $\beta$-model in four
one-dimensional parameterizations. These two models differ with how the original data
with asymmetric uncertainties are treated. The upper two panels use the original
spherical $\da$ set and the lower two employ the $\da$ set, which has been shifted in
terms of the method of \citet{d'a04}. The left two panels correspond to method (I) while
the right two method (II).}
\end{figure*}

\begin{figure*}\label{figure4}
\epsscale{1.15} \plottwo{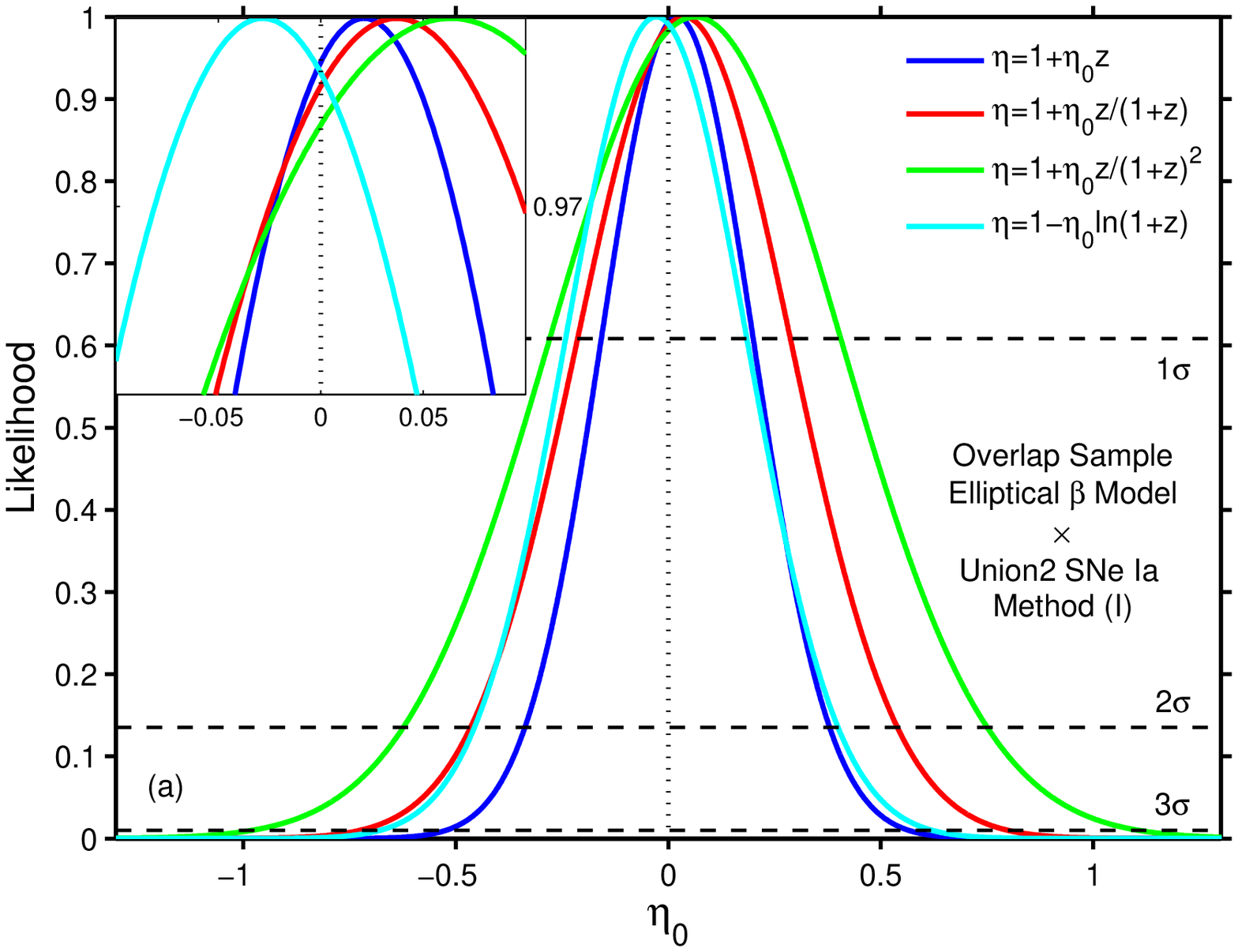}{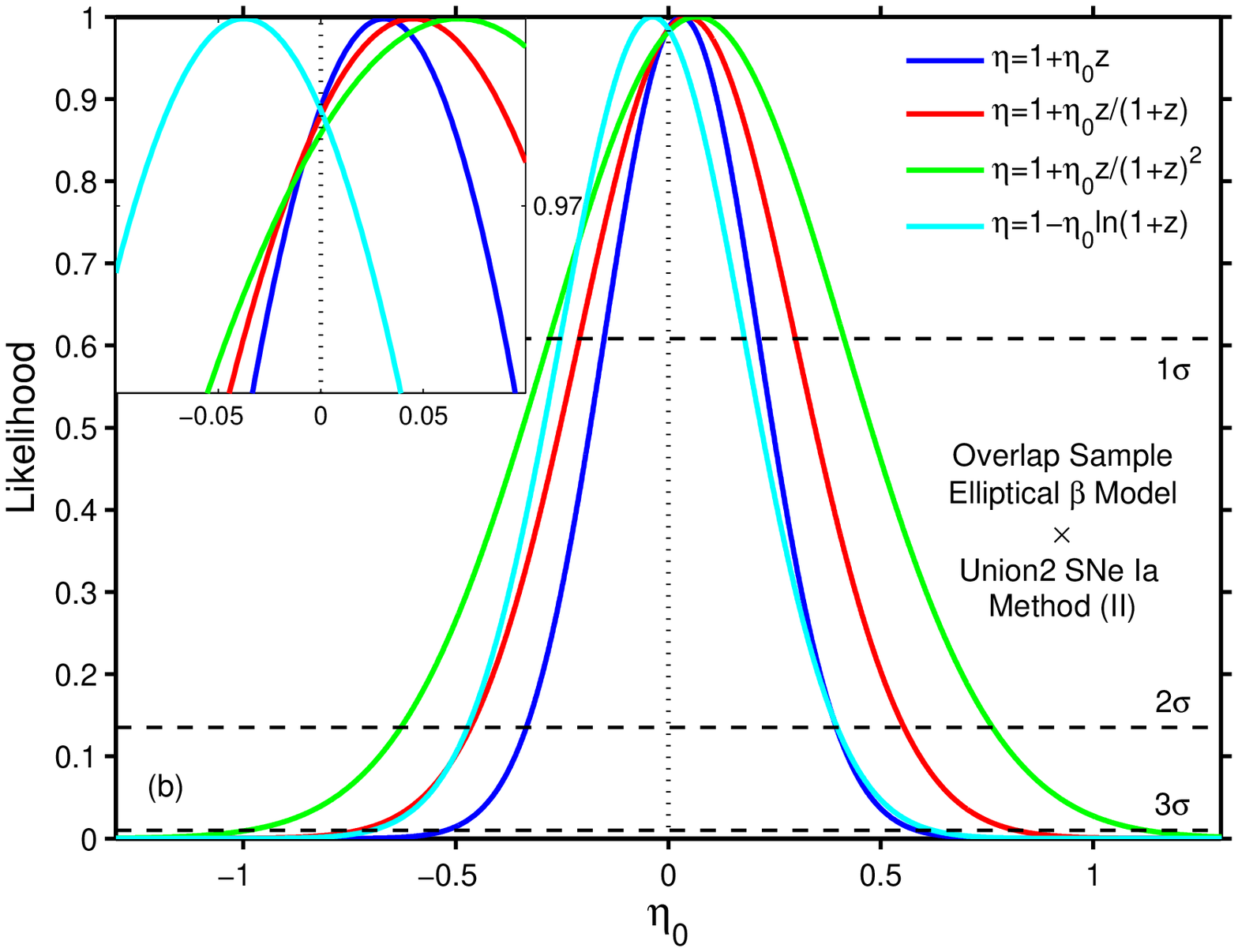} \caption{Likelihood distribution functions
for the overlap sample under elliptical $\beta$-model in four one-dimensional
parameterizations. Panel (a) corresponds to the results of SNe Ia data obtained with
method (I), and panel (b) corresponds to those with method (II). The inserts of panel (a)
and (b) give zoom-in views near the peaks of the likelihood functions.}
\end{figure*}

\begin{figure*}\label{figure5}
\epsscale{1.15} \plottwo{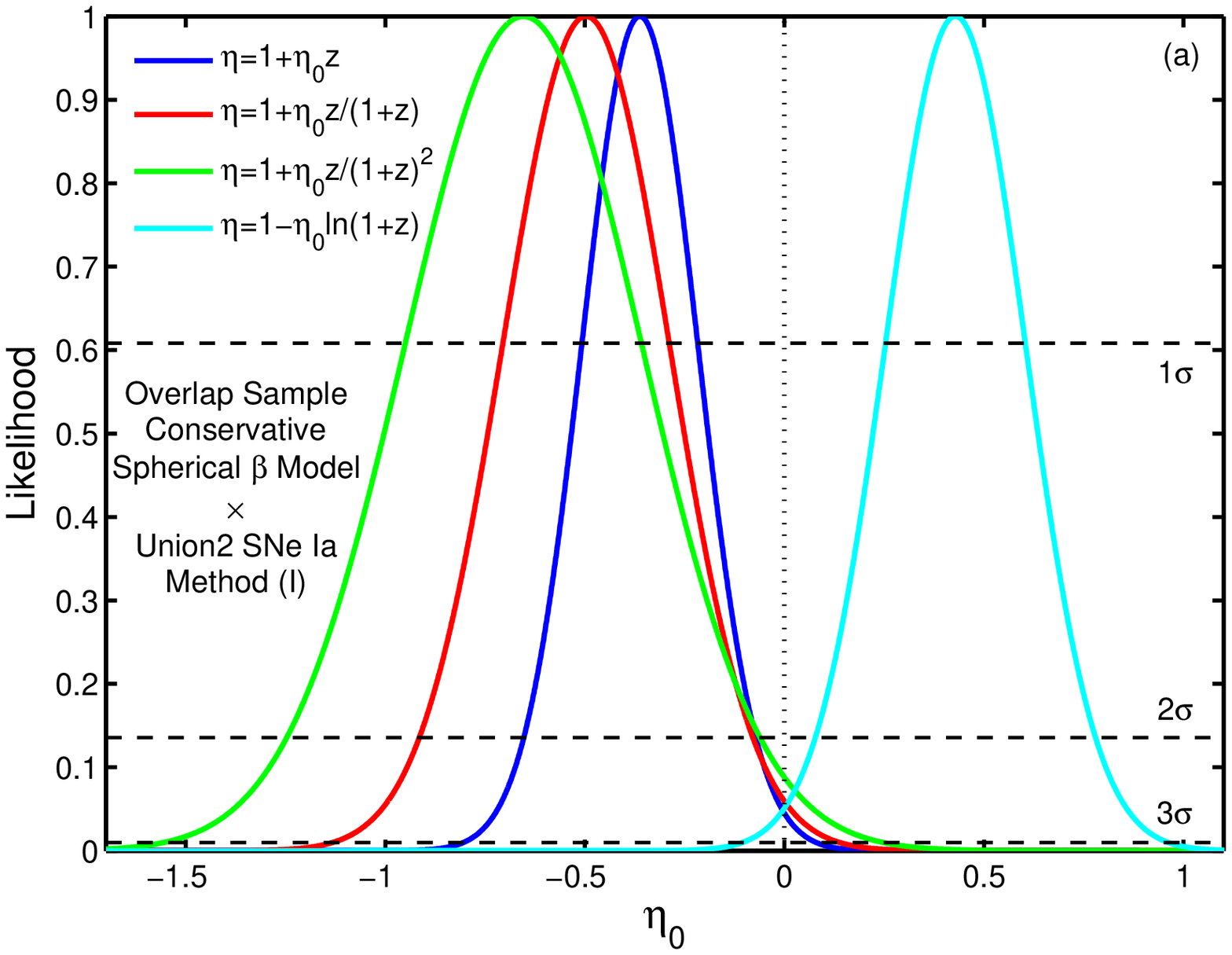}{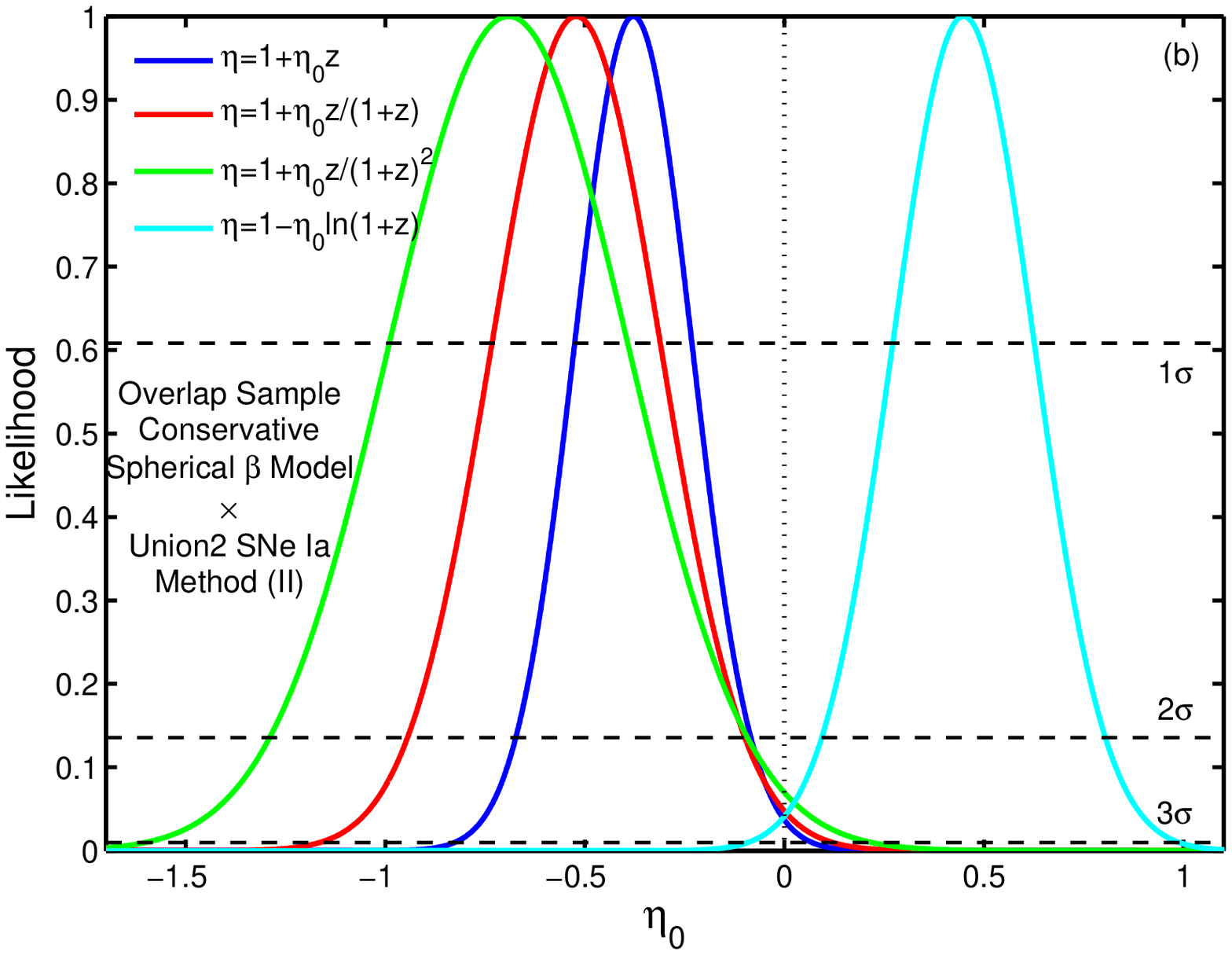} \epsscale{1.15}
\plottwo{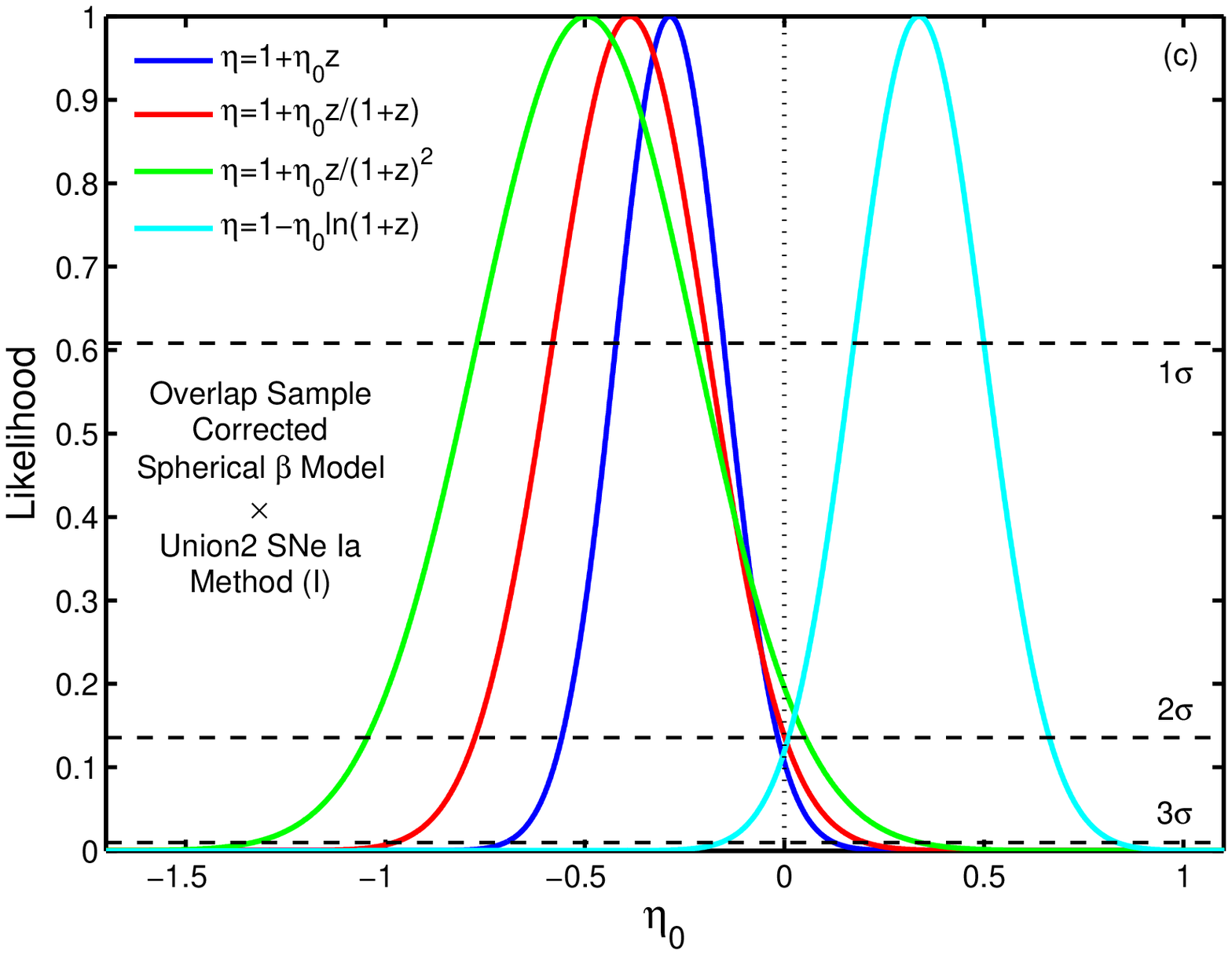}{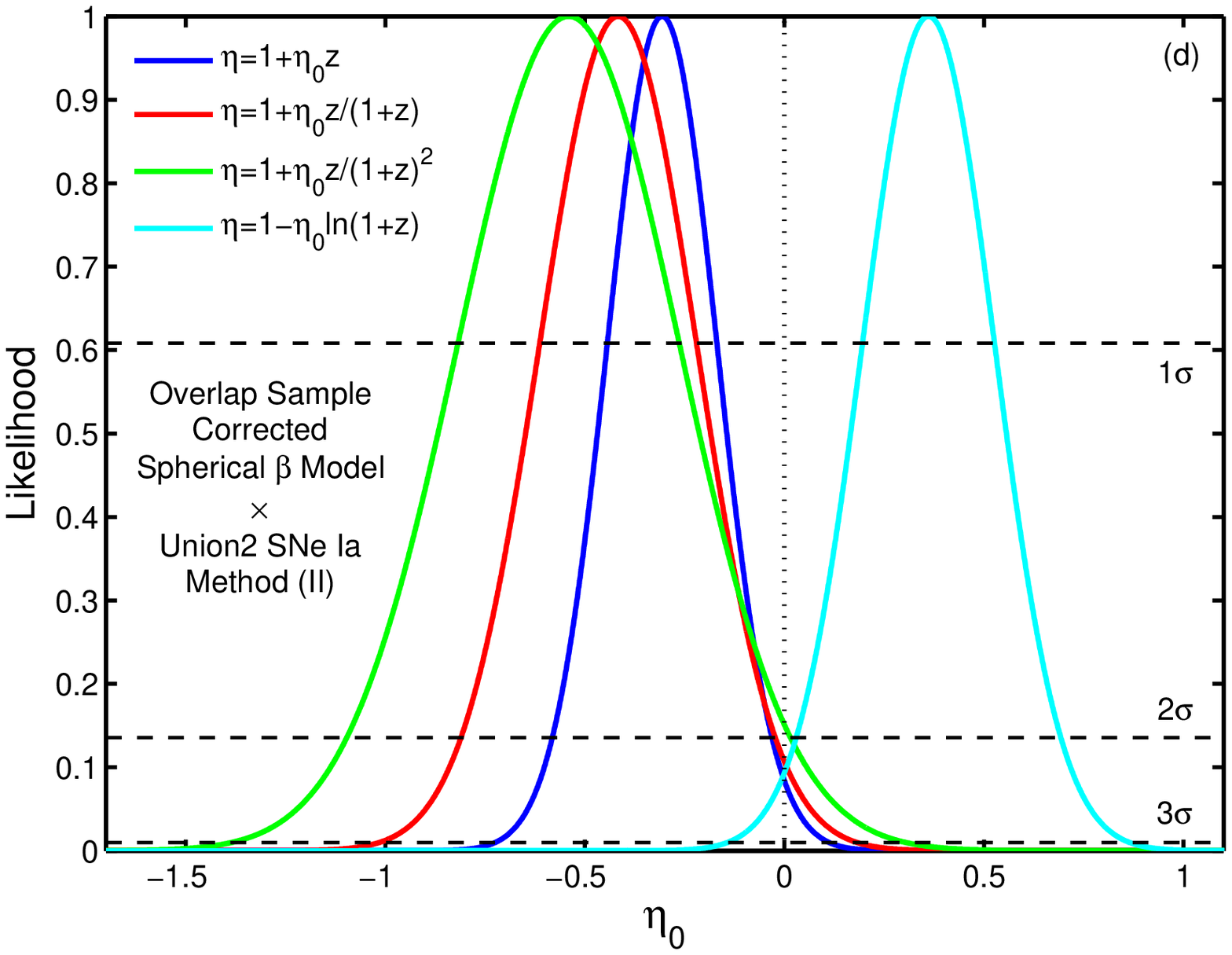} \caption{Likelihood distribution functions for the overlap
sample under conservative spherical $\beta$-model and corrected spherical $\beta$-model
in four one-dimensional parameterizations. These two models differ with how the original
data with asymmetric uncertainties are treated. The upper two panels use the original
spherical $\da$ set the and the lower two employ the $\da$ set, which has been shifted in
terms of the method of \citet{d'a04}. The left two panels correspond to method (I) while
the right two method (II).}
\end{figure*}

It is straightforward to introduce the test parameter $\eta$ according to
Eq.~(\ref{eq:da-dl}),
\begin{equation}\label{eq:dadleta}
\frac{\dl(z)}{\da(z)(1+z)^2}=\eta.
\end{equation}
Then, the $\eta_0$-associated one-dimensional parameterizations for $\eta$ are expressed
as follows,
\begin{mathletters}
\begin{eqnarray}
\eta(z)&=&1+\eta_0z ,\label{eq:eta0_1}\\
\eta(z)&=&1+\eta_0z/(1+z) ,\\
\eta(z)&=&1+\eta_0z/(1+z)^2 ,\\
\eta(z)&=&1-\eta_0\rm{ln}(1+z).\label{eq:eta0_4}
\end{eqnarray}
\end{mathletters}
Note that $\da(z)$ can not be obtained directly from the techniques of SZE plus X-ray
surface brightness observations. \citet{uzan04} gave a relation that describes the
angular diameter distance determined from observations, i.e. $\da(z)=\dac(z)\eta^{-2}$,
which reduces to $\da(z)$ only when $\eta=1$, i.e. the DD relation holds with no
violation. Combining this relation with Eq.~(\ref{eq:dadleta}), it is straightforward to
achieve the target equation for further statistical analysis, i.e.
\begin{equation}\label{eq:dacdl}
\frac{\dac(z)\left( 1 + z \right)^2}{\dl(z)} = \eta,
\end{equation}
where the expression of $\eta$ is checked four times according to
Eqs.~(\ref{eq:eta0_1})--(\ref{eq:eta0_4}). Maximum likelihood estimation is employed to
determine the most probable values for the parameters, via $\mathbb{L} \propto
\mathrm{e}^{-\chi^2/2}$ and
\begin{equation}\label{eq:chi2}
\chi^2 = \sum\limits_z {\frac{\left( \eta(z)-\eta_{\rm obs}(z)
\right)^2}{\sigma_{\eta_{\rm obs}}^2}},
\end{equation}
where $\eta_{\rm obs}(z) = \left( 1 + z \right)^2 \dac(z)/\dl(z)$ and its uncertainty
$\sigma_{\eta_{\rm obs}}$ is given by
\begin{equation}\label{eq:sigeta}
\sigma_{\eta_{\rm obs}}^2 = {\eta_{\rm obs}^2}\left[\left(
\frac{\sigma_{\dac(z)}}{\dac(z)} \right)^2 + \left(\frac{\sigma
_{\dl(z)}}{\dl(z)}\right)^2 \right].
\end{equation}

Utilizing methods (I) and (II) described in Section 2, we are able to determine all
corresponding $\dl$ data at each cluster's redshift. Moreover, conservative spherical
$\beta$-model and corrected spherical $\beta$-model are retrieved from the original data
of the $r<100$kpc-cut isothermal $\beta$-model reported by \citet{bona06} (see Table 1).
In Table 1, the number of SNe Ia selected for each galaxy cluster for method (II)
analyses is also listed. For the majority of galaxy clusters, more than one SNe Ia are
used. Thus the statistics is improved, especially for nearby cluster sample, e.g.
Abell~1656. Because there is no SN satisfying $|\Delta z|<0.005$ from the cluster CL
J1226.9+3332 (the nearest one is at $\Delta z=0.005$), this object is excluded, and also
removed from method (I) analyses for the sake of direct comparison between our two
methods. Our methods not only avoid double counting of any SNe Ia data, but also take
into account all possible clusters data.

According to the results shown in Figure 2 and Table 2, the consistency is clear between
the DD relation and the elliptical geometry hypothesis, since the DD relation value
($\eta_0=0$) is always within $1\sigma$ CL of the best-fit values of elliptical
$\beta$-model using both methods. However the best-fit values of conservative and
corrected spherical $\beta$-models depart from $\eta_0=0$ at nearly $3\sigma$ CL (see
Figure 3), especially for conservative spherical $\beta$-model (both methods) and
corrected spherical $\beta$-model (method (II)).

As to the overlap sample, the difference between the two morphological models are more
prominent. Figure 4 demonstrates good consistency between elliptical $\beta$-model and
the DD relation, with the best-fit values for all parameterizations close to the DD
relation value ($\eta_0=0$). For the two spherical models, the results in Figure 5 show a
rather marginal compatibility with the DD relation, which is slightly worse than the
whole spherical sample results (see Figure 3). Statistical results of each sample using
the two methods are listed in Table 2 for comparison.

The one-dimensional parameterizations of the deviation from the DD relation, i.e. $\eta$
in Eqs.~(\ref{eq:eta0_1})--(\ref{eq:eta0_4}), may be somewhat restrictive in terms of the
degrees of freedom. Here the DD relation is tested with two-dimensional parameterizations
of $\eta$,
\begin{mathletters}
\begin{eqnarray}\label{eta12}
\eta(z)&=&\eta_1+\eta_2z ,\\
\eta(z)&=&\eta_1+\eta_2z/(1+z) ,\\
\eta(z)&=&\eta_1+\eta_2z/(1+z)^2 ,\\
\eta(z)&=&\eta_1-\eta_2\rm{ln}(1+z) ,
\end{eqnarray}
\end{mathletters}
where ($\eta_1,\eta_2$) has the value of (1,0) if the DD relation holds. It is found that
(1,0) falls in the $1\sigma$ region of all two-dimensional parameterizations for
elliptical $\beta$-model (Figure 6), regardless of the method used. However, for the
conservative and corrected spherical $\beta$-models, the DD relation can only be
accommodated at $3\sigma$ CL for both methods, except the result ($2\sigma$) with method
(II) for the corrected spherical $\beta$-model.

The results of the two-dimensional analysis of the overlap sample (see Figures 8 and 9)
are in good agreement with those of its one-dimensional analysis. In fact, the overlap
sample shows a stronger preference for the elliptical geometry, as the DD relation is
almost always $3\sigma$ away from the best-fit parameters for the spherical samples.

Table 3 presents the minimum values of reduced chi square ($\chi^2_{\rm min}/{\rm
d.o.f.}$) for each models with the two methods. Generally speaking, it shows the fitting
quality of the corresponding models to each sample of galaxy clusters. Consistent with
the results given by the best-fit values of $\eta_0$ in Table~2, the results of
$\chi^2_{\rm min}/{\rm d.o.f.}$ also favor elliptical model given the validity of DD
relation, since $\chi^2_{\rm min}/{\rm d.o.f.}$ of this model is much closer to unity
than those of spherical models.

In sum, not only does spherical models have larger departures of the best-fit values of
$\eta_0$ from the DD relation value ($\eta_0=0$) as shown in Table~2, spherical models
also provide poorer fits than elliptical models do as presented in Table~3. In
consequence, we argue that the galaxy cluster sample modeled by ellipsoidal morphology is
a better understanding of cluster's intrinsic nature than those modeled by spherical
morphology.

\begin{figure*}\label{figure6}
\epsscale{1.15} \plottwo{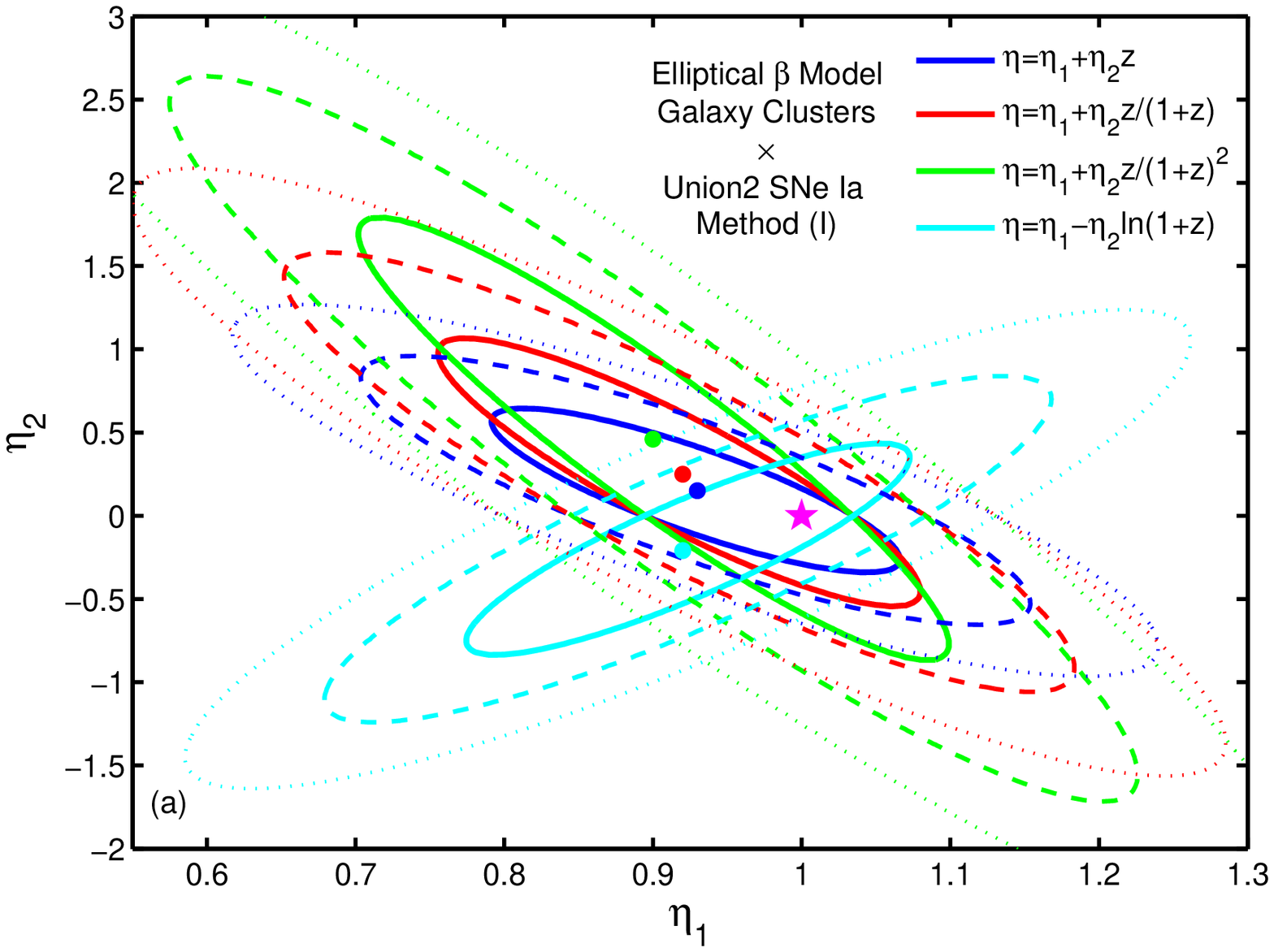}{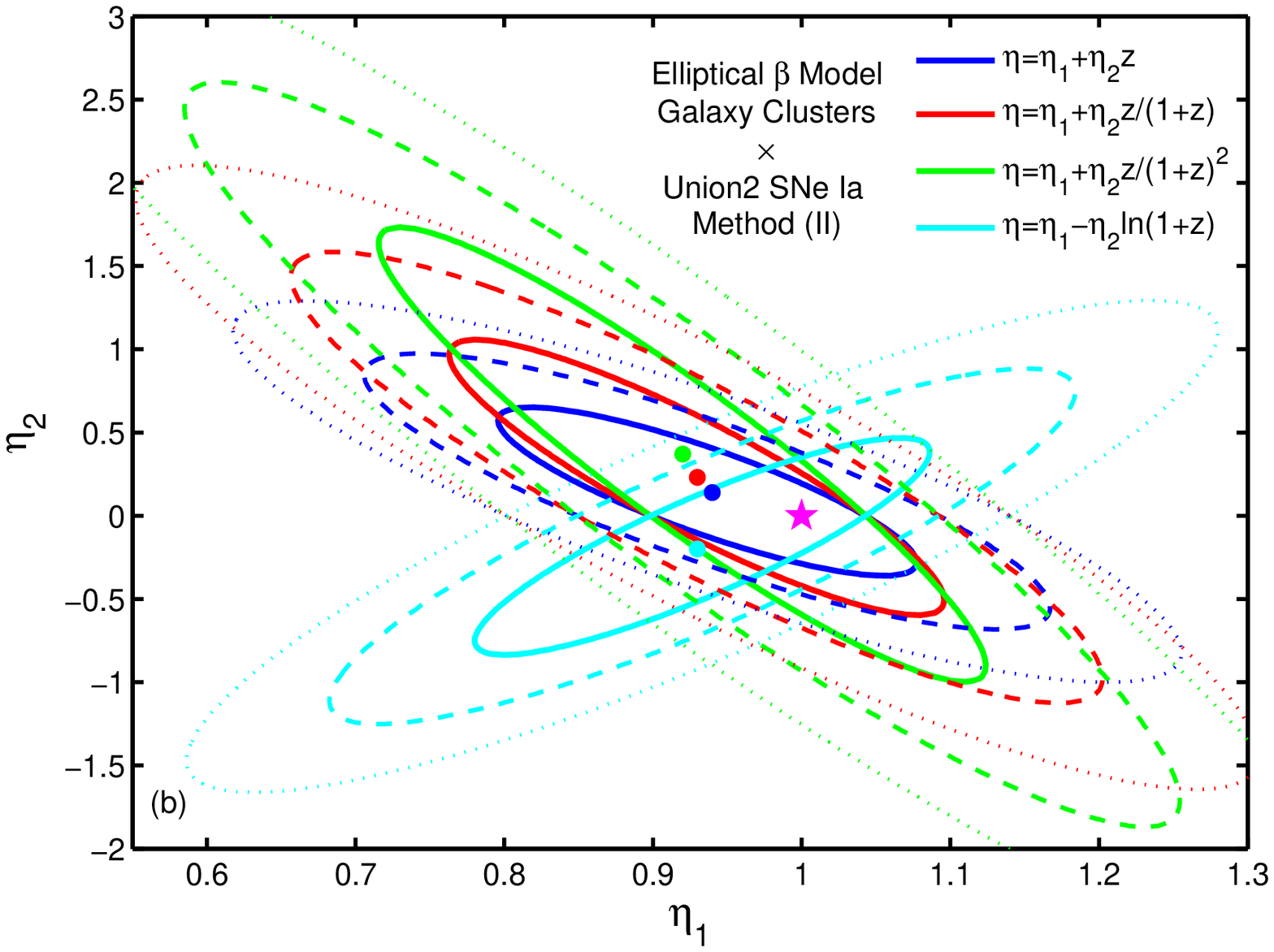} \caption{Likelihood distribution functions
for elliptical $\beta$-model in four two-dimensional parameterizations. As in Figure 2,
panel (a) corresponds to method (I) while panel (b) method (II). The 1-, 2- and
3-$\sigma$ CLs are plotted by solid, dashed and dotted lines respectively. The pentagram
in each panel stands for the DD relation value (1,0), while the big dots in corresponding
colors represent the actual best-fit values for four parameterizations.}
\end{figure*}

\begin{figure*}\label{figure7}
\epsscale{1.15} \plottwo{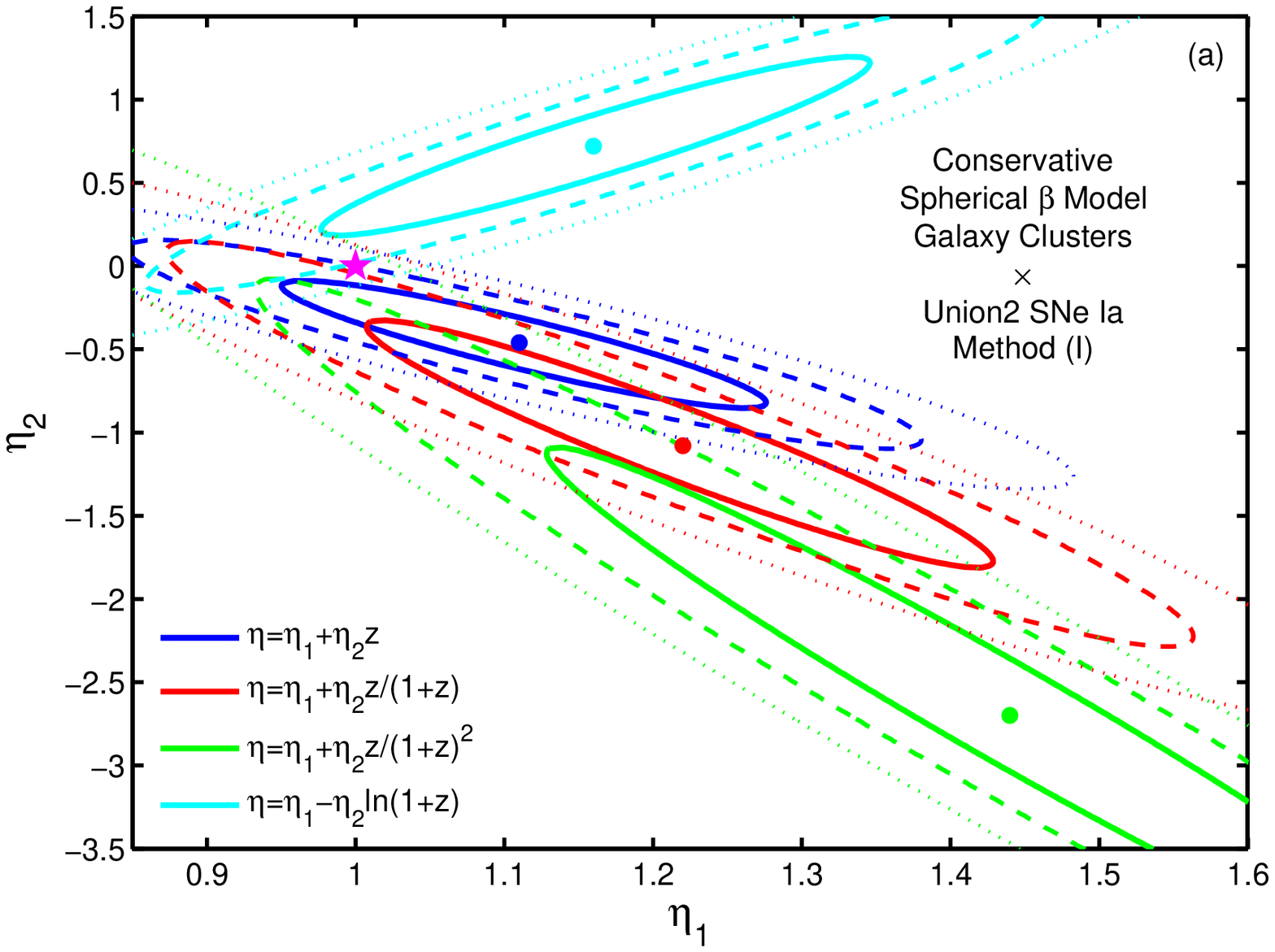}{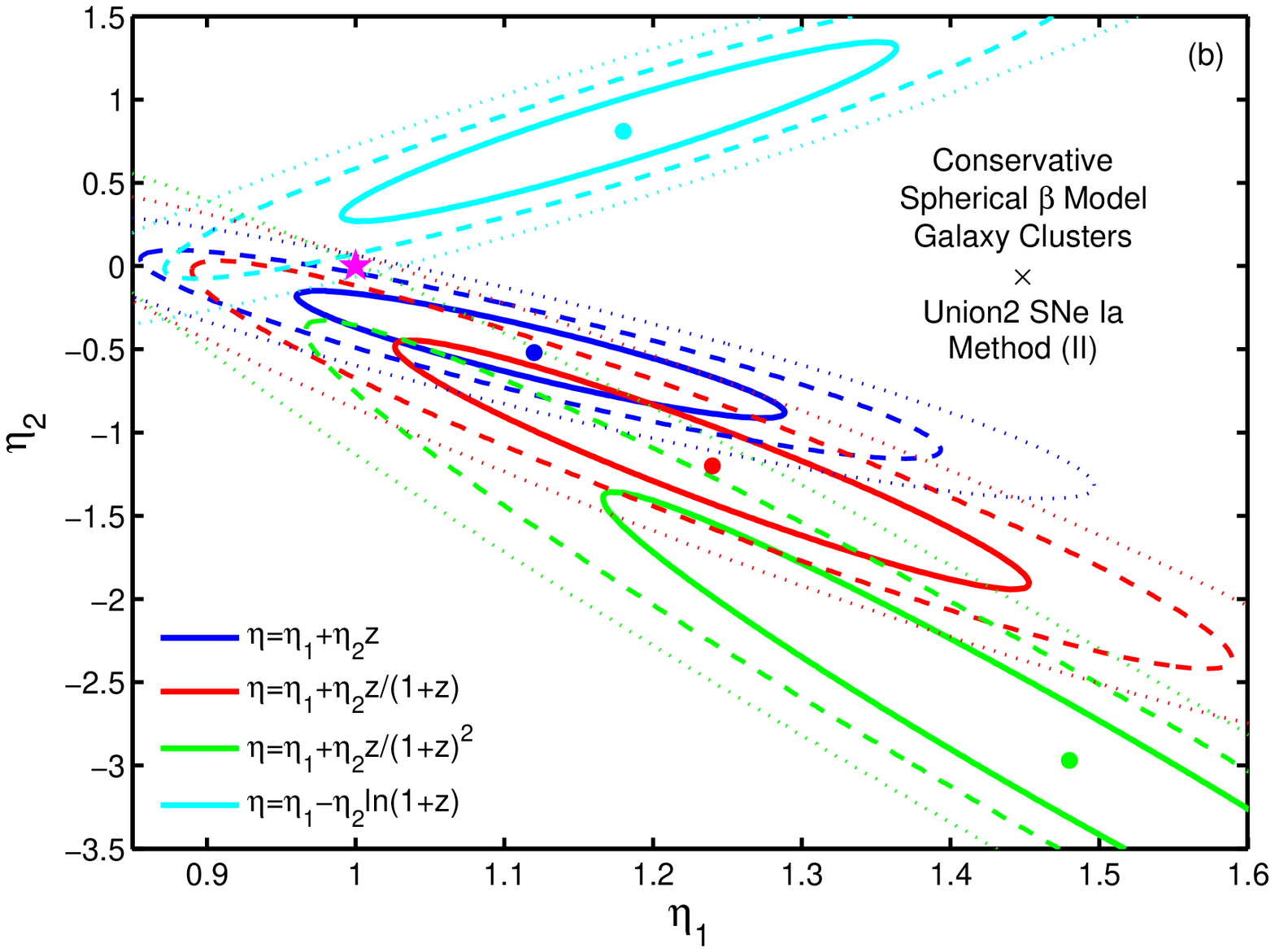} \epsscale{1.15}
\plottwo{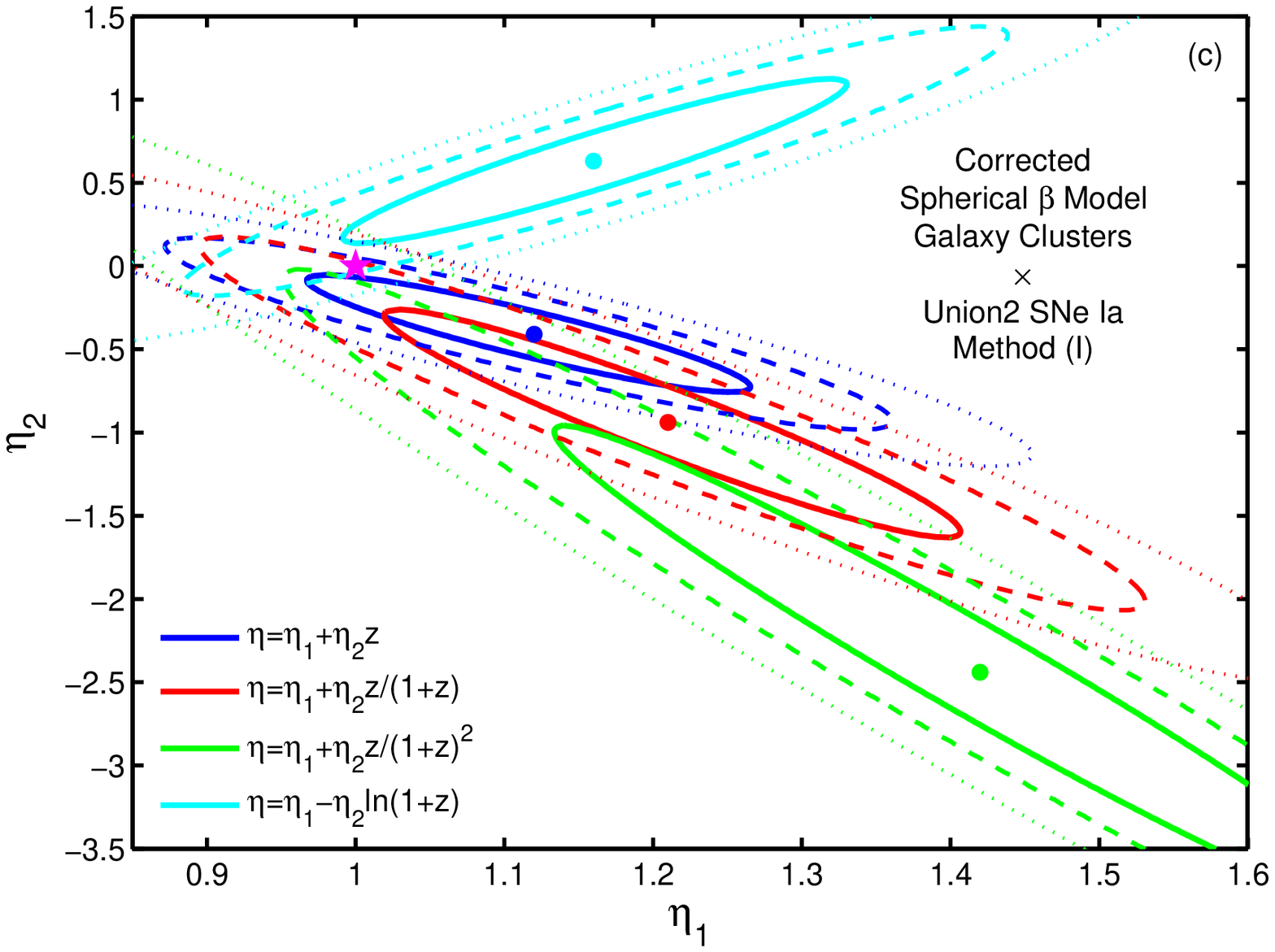}{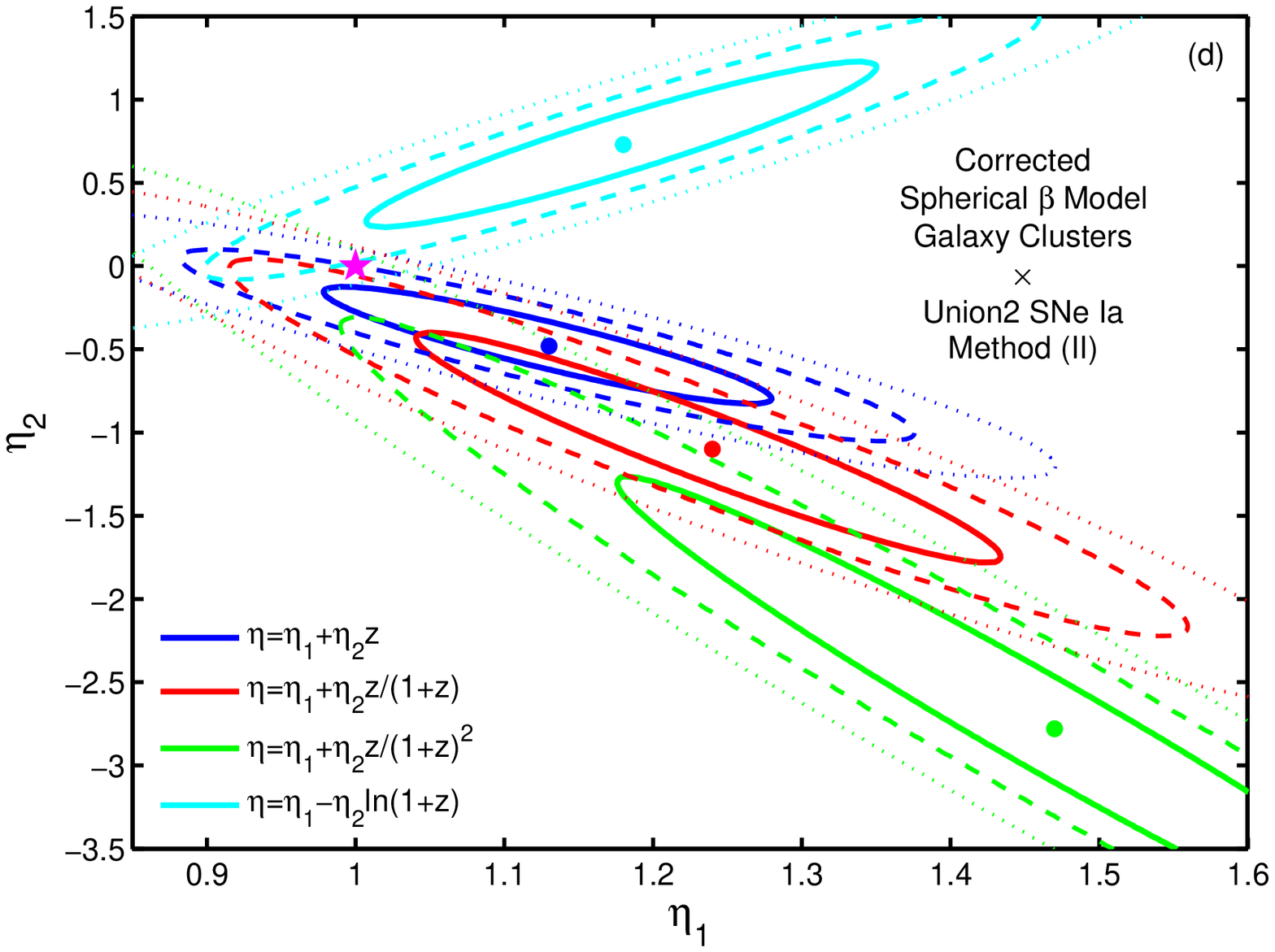} \caption{Likelihood distribution functions for
conservative spherical $\beta$-model and corrected spherical $\beta$-model in four
two-dimensional parameterizations. As in Figure 3, the upper two panels are from
conservative model with lower two corrected model. The left two panels are obtained
through method (I) while right two method (II). The 1-, 2- and 3-$\sigma$ CLs are plotted
by solid, dashed and dotted lines respectively. The pentagram in each panel stands for
the DD relation value (1,0) while the big dots in corresponding colors represent the
actual best-fit values for four parameterizations.}
\end{figure*}

\begin{figure*}\label{figure8}
\epsscale{1.15} \plottwo{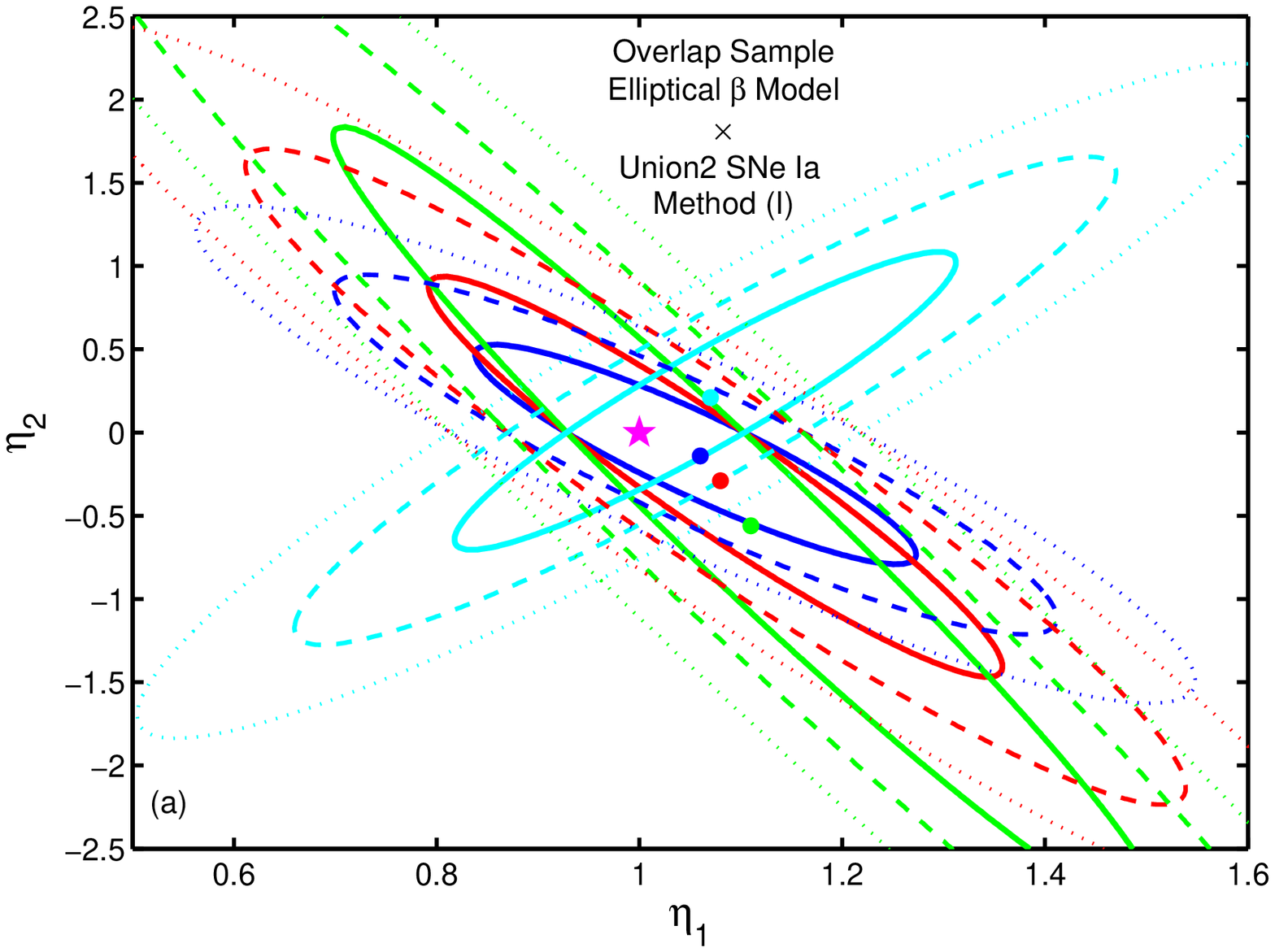}{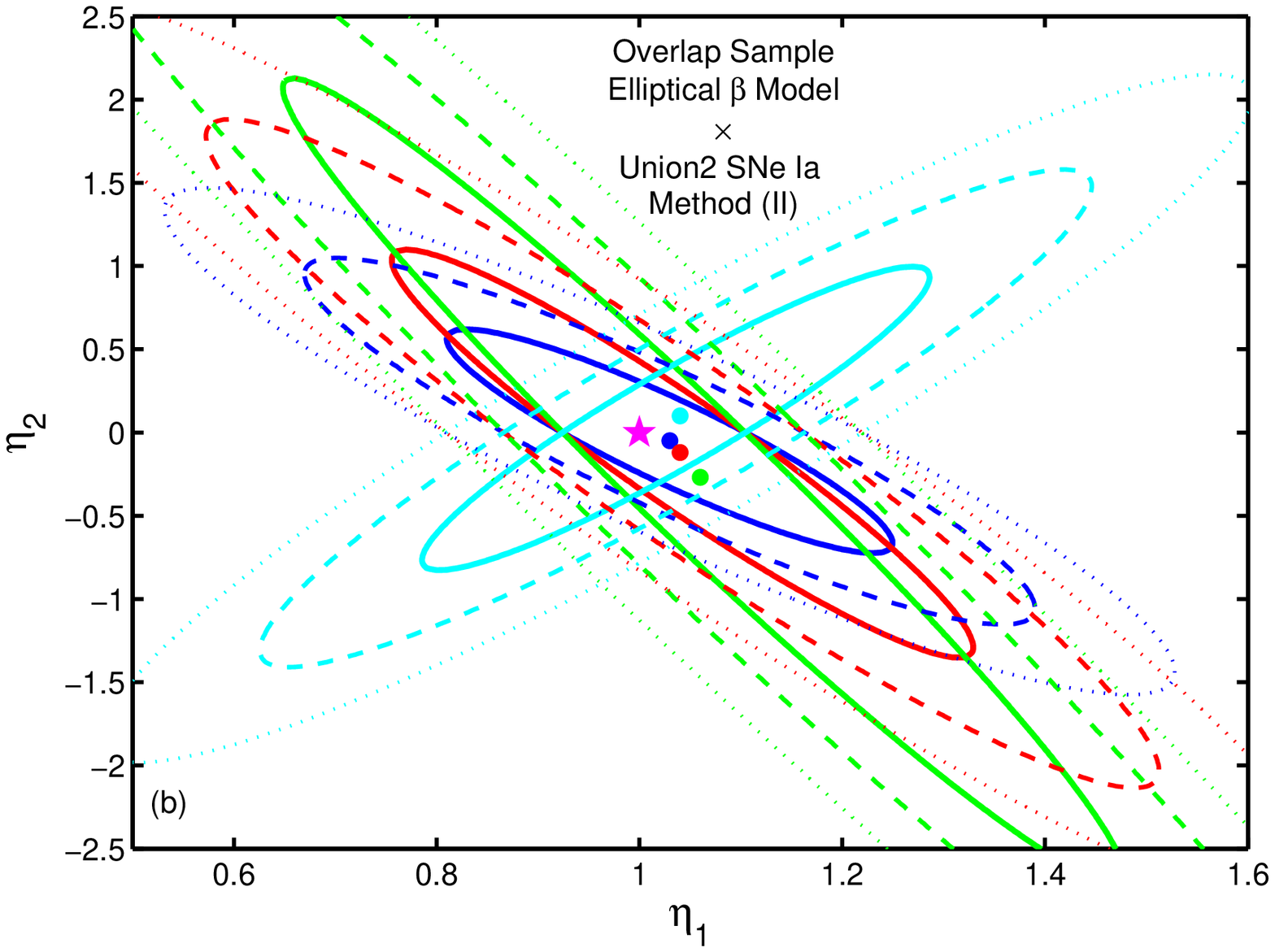} \caption{Likelihood distribution functions
for the overlap sample under elliptical $\beta$-model in four two-dimensional
parameterizations. As in Figure 4, panel (a) corresponds to method (I) while panel (b)
method (II). Different colors of confidence contours refer to different parameterizations
as shown in Figure 6. The 1-, 2- and 3-$\sigma$ CLs are plotted by solid, dashed and
dotted lines respectively. The pentagram in each panel stands for the DD relation value
(1,0), while the big dots in corresponding colors represent the actual best-fit values
for four parameterizations.}
\end{figure*}

\begin{figure*}\label{figure9}
\epsscale{1.15} \plottwo{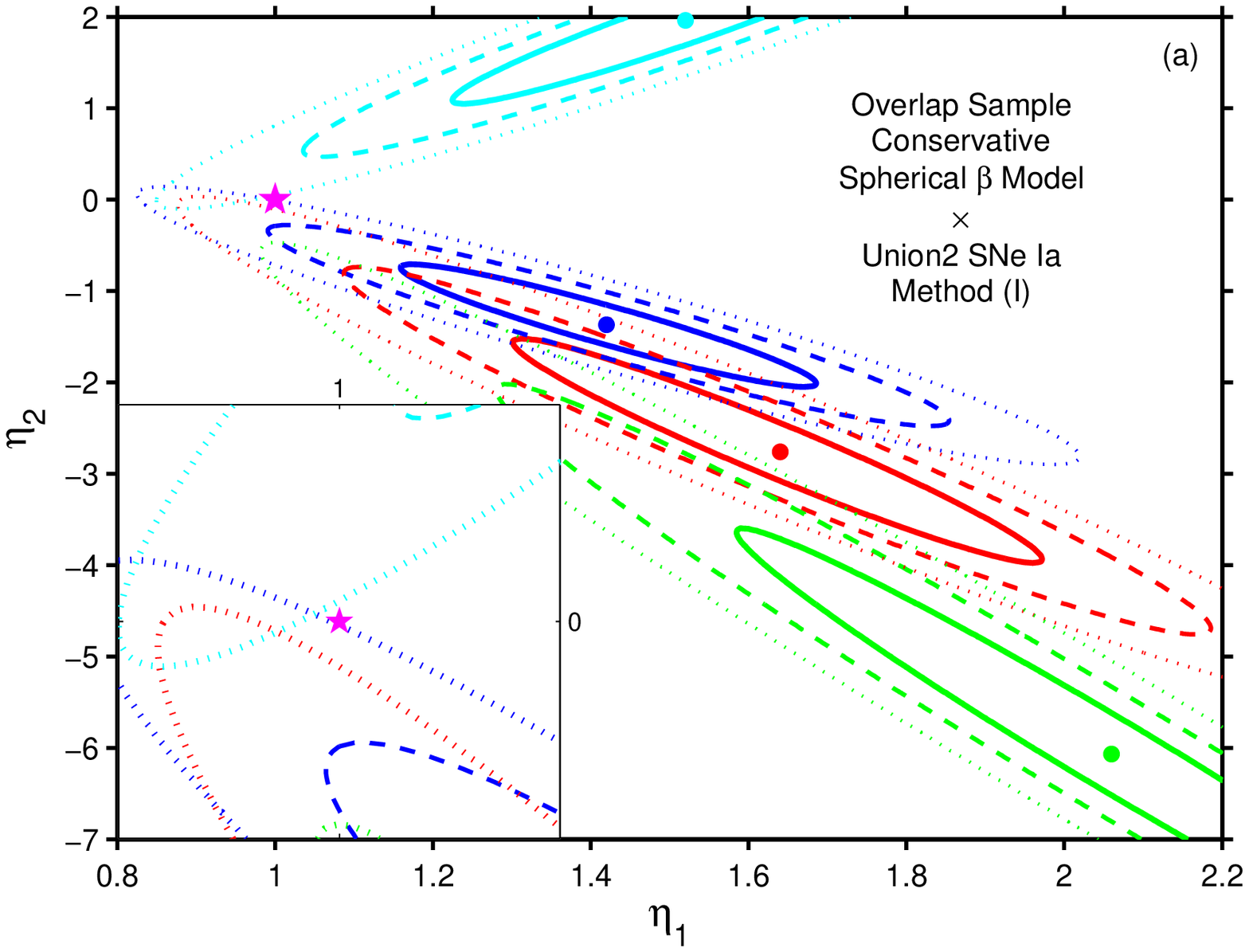}{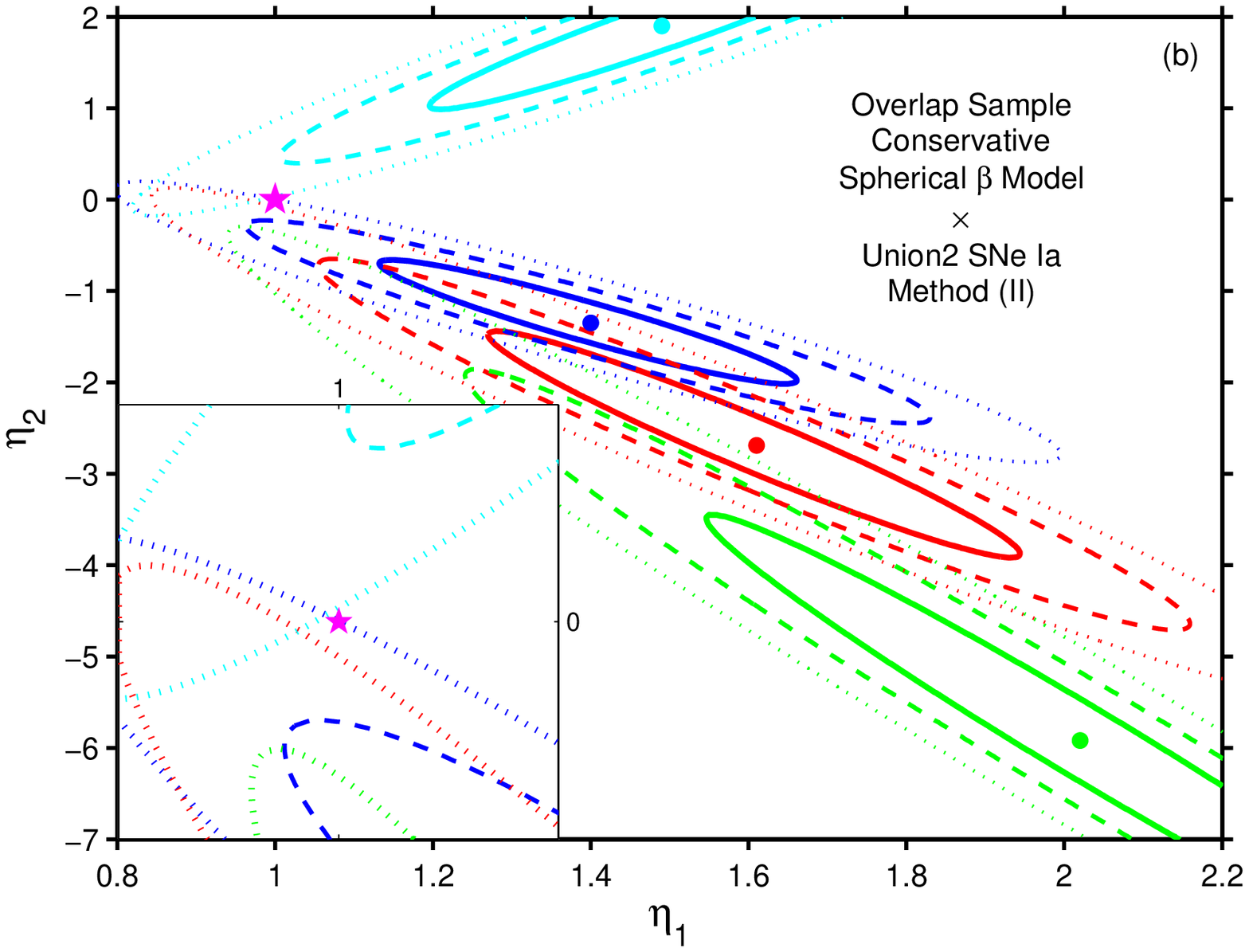} \epsscale{1.15}
\plottwo{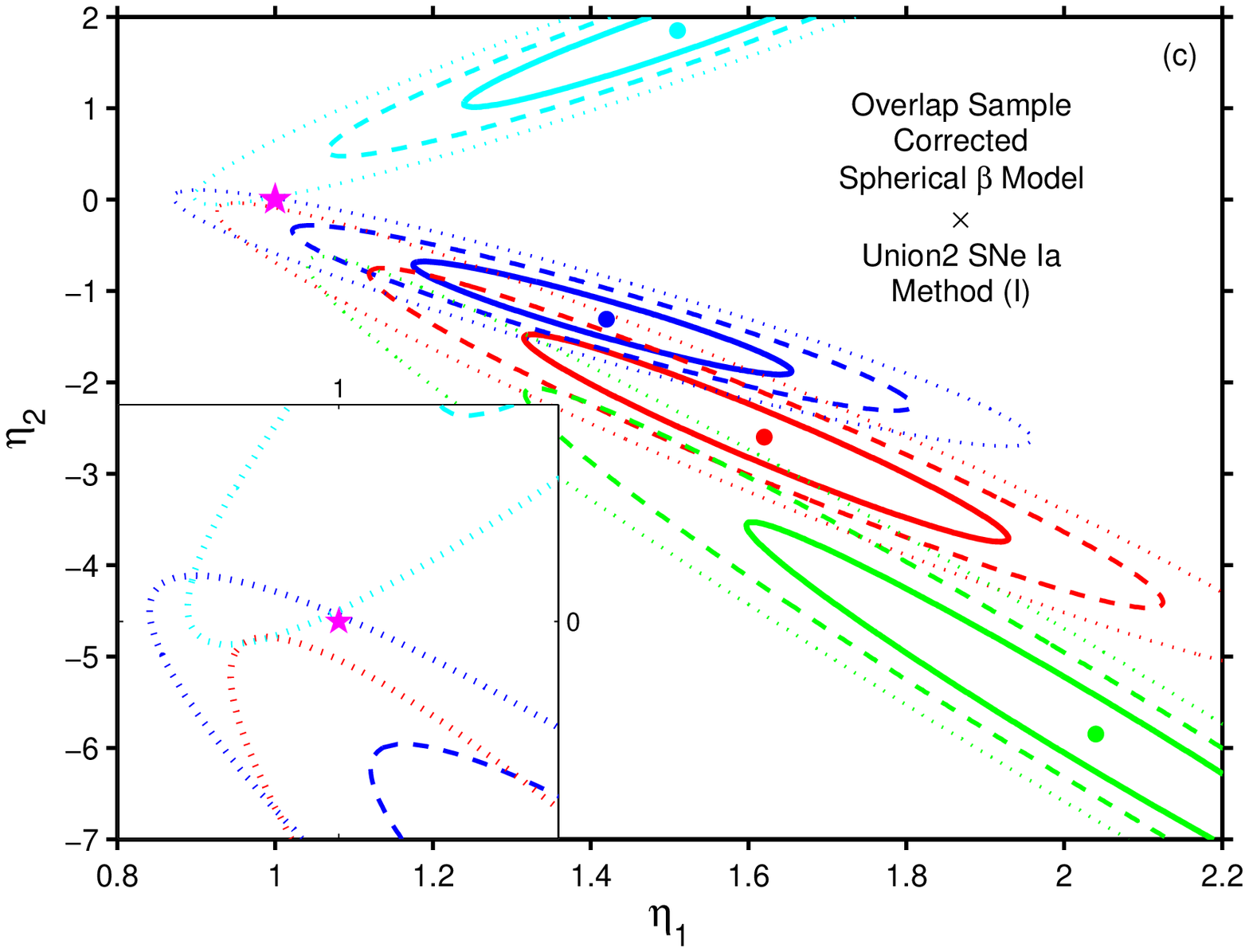}{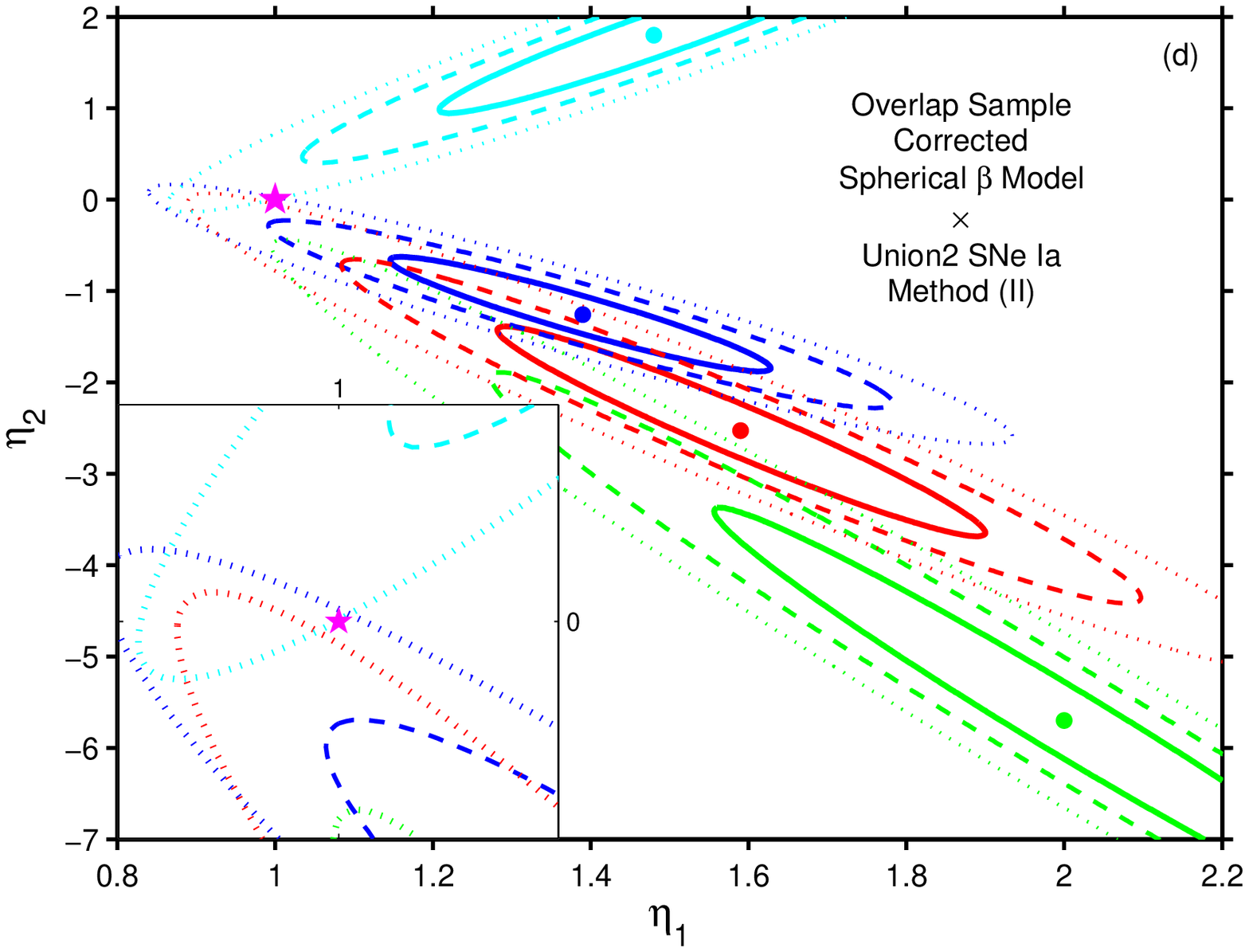} \caption{Likelihood distribution functions for the overlap
sample under conservative spherical $\beta$-model and corrected spherical $\beta$-model
in four two-dimensional parameterizations. As in Figure 5, the upper two panels are from
conservative model with lower two corrected model. The left two panels are obtained
through method (I) while right two method (II). Different colors of confidence contours
refer to different parameterizations as shown in Figure 7. The 1-, 2- and 3-$\sigma$ CLs
are plotted by solid, dashed and dotted lines respectively. The pentagram in each panel
stands for the the DD relation value (1,0), while the big dots in corresponding colors
represent the actual best-fit values for four parameterizations. The inserts of four
panels give zoom-in views near the the DD relation value (1,0), which is around the edges
of 3-$\sigma$ CL contours of all likelihood distributions.}
\end{figure*}

\section{Conclusions and Discussions}

In this paper, motivated by the investigation on intrinsic structure of galaxy clusters,
we have tested the validity of the DD relation using luminosity distances from Union2 SNe
Ia and angular diameter distances from two cluster morphological models, namely, triaxial
ellipsoidal $\beta$-model and spherical $\beta$-model. In order to obtain more reliable
results, two sub-samples of spherical $\beta$-model are analyzed according to different
treatments of the two-sided errors of $\da$ data, i.e. conservative and corrected
spherical $\beta$-models. Moreover, in order to directly compare these two types of
morphological models as well as minimize systematic uncertainties of the test, we also
conduct the analysis on the overlap sample, i.e. the same set of cluster individuals
under both ellipsoidal and spherical $\beta$-models for electron density distribution
profiles.

To test the DD relation, it is assumed that $\eta(z)=\dac(z)(1+z)^2/\dl(z)$. In practice,
the reduction of statistical errors is realized by two methods, (I) fitting the Union2
SNe Ia data with weighted least-squares and interpolating $\dl$ at each cluster's
redshift, (II) binning the SNe Ia data within the redshift range $|z_{\rm cluster} -
z_{\rm SN}|<0.005$ to get $\dl$ at the cluster's redshift. In methods (I) and (II), the
$\eta(z)$ parameter is parameterized in four one-dimensional forms,
$\eta(z)=1+\eta_0z,~\eta(z)=1+\eta_0z/(1+z),~\eta(z)=1+\eta_0z/(1+z)^2~\textrm{and}~\eta(z)=1-\eta_0\rm{ln}(1+z)$.
Other four more general parameterizations are also considered, i.e.
$\eta(z)=\eta_1+\eta_2z, ~\eta(z)=\eta_1+\eta_2z/(1+z), ~\eta(z)=\eta_1+\eta_2z/(1+z)^2
~\textrm{and} ~\eta(z)=\eta_1-\eta_2\rm{ln}(1+z)$, which are designated to describe the
two-dimensional admissible parameter space \citep{2011JCAP...05..023N}.

Maximum likelihood analysis is used to fit the parameters, $\eta_0$ or ($\eta_1$,
$\eta_2$). Our results show that for elliptical $\beta$-model, regardless of whether
method (I) or (II) is employed, the DD relation values ($\eta_0=0$ or
$\eta_1=1,\eta_2=0$) are always in $1\sigma$ region for all one- and two-dimensional
forms. The results of elliptical $\beta$-model support the idea that elliptical geometry
is more consistent with non-violation of the DD relation. In the case of conservative
spherical $\beta$-model, with both methods, it is found that the DD relation value is
only marginally consistent at $3\sigma$ with the best-fit values via one- and
two-dimensional analyses. The result from corrected spherical $\beta$-model is that the
DD relation can be accommodated at $3\sigma$ CL for one-dimensional forms except using
method (I), and similar results is obtained using both methods for two-dimensional
parameterizations. In consequence, we can not prove that spherical $\beta$-model is
compatible with the validity of the DD relation.

Nevertheless, it is noticed that these results may be to some extent significantly
affected by some individual data points. For instance, if cluster CL J1226.9+3332 is
still included in method (I) (this object is naturally excluded in method (II) for there
is no SN Ia satisfying $\Delta z<0.005$), the conclusion will be transformed into that
the DD relation cannot be accommodated even at 3$\sigma$ CL for conservative spherical
model. This result is tremendously different from the original results shown in
Figures~3(a) and 7(a). Another example is cluster RX J1347.5-1145. If we exclude this
cluster from the overlap sample, and apply our method to this new sample, the results
show that for conservative spherical model the DD relation can be accommodated at
2$\sigma$ CL, and its $\chi^2_{\rm min}/{\rm d. o. f.}$ give 22.470/15 and 20.204/14 for
one- and two-dimensional parameterizations respectively. This result is not similar to
our original results neither. Thus some thorough investigations on the influence of every
cluster on the global compatibility should be paid attention to in future work.

Furthermore, it is worth making a comparison between the two methods. Our results
demonstrate that the DD relation is compatible with elliptical $\beta$-model at $1\sigma$
not only for one-dimensional parameterizations but also for two-dimensional forms,
regardless of the methods. There is no discrepancy between the two methods for
conservative spherical $\beta$-model neither; the DD relation can be accommodated at
$3\sigma$ with the two methods. Likewise, corrected spherical $\beta$-model has
approximately similar outcome, using the fitting and binning methods. The results of
$\chi^2_{\rm min}/{\rm d. o. f.}$ are similar between the two methods, as presented by
Table~3. Accordingly, the two methods we used are self-consistent. With our analysis, the
marked triaxial ellipsoidal model is a more reasonable hypothesis describing the
structure of the galaxy cluster compared with the spherical hypothesis if the DD relation
is valid in cosmological observations.

\citet{bona06} discussed the sources of some statistical and systematic uncertainties
capable of affecting the angular diameter distance measurements. The main origins of
systematic errors are calibrations of SZE, X-ray absolute flux and temperature, while the
major statistical uncertainties come from SZE observations, X-ray spectra and images.
Interestingly \citet{bona06} also pointed out that the statistical uncertainty by cluster
asphericity has the largest effect, i.e. 15\% on $\da$ \citep[see Table 3 of][]{bona06},
but will be averaged out when a large ensemble of clusters is considered
\citep{1999ApJ...522...59S}. However, our main results tell that in describing the
three-dimensional morphology of galaxy clusters, the elliptical $\beta$-model prevails
over the spherical $\beta$-models. We argue that the current cluster sample with well
measured $\da$ is not sufficient large to average out the uncertainty by cluster
asphericity. Actually, several up-to-date observational work strongly support the
triaxial ellipsoidal morphology for galaxy clusters, e.g. observations in X-ray and
strong gravitational-lensing \citep{2010ApJ...713..491M}, SZE imaging
\citep{2011ApJ...728...39S}, etc.. Recently \citet{2007ApJ...668....1N} proposed another
candidate profile for the distribution of intracluster medium, known as Nagai model,
which is based on the well known NFW dark matter profile
\citep{1996ApJ...462..563N,1997ApJ...490..493N}. Although analytically non-integrable,
this model can also reproduce observed intracluster medium properties through
simulations. However recent observation is incapable of distinguishing between Nagai
model and elliptical $\beta$-model, due to limited spatial resolution
\citep{2011ApJ...728...39S}. Maybe with more advanced detection techniques in the future,
people can learn more about the intrinsic morphology of galaxy clusters from direct
observations.

\acknowledgments \noindent \emph{Acknowledgments}. We are greatly indebted to the
anonymous referee for insightful suggestions that enable us to develop a more
comprehensive study. Prof. Charling Tao is deeply acknowledged for her patient and
penetrating instruction on data analysis of SNe Ia. We also cordially thank Prof. C.-P.
Yuan and Qiao Wang for reading of this manuscript. XL M is grateful to Cong Ma for useful
discussion also. This work was supported by the National Science Foundation of China
(Grant No. 11173006), the Ministry of Science and Technology National Basic Science
program (project 973) under grant No. 2012CB821804, the Fundamental Research Funds for
the Central Universities, and the Bairen program from the Chinese Academy of Sciences and
the National Basic Research Program of China grant No. 2010CB833000.



\begin{deluxetable}{llrrrrrrr}
\tablecolumns{9}
\tabletypesize{\scriptsize} \tablewidth{0pt} \tablecaption{Galaxy Cluster Data}
\label{table:gc}
\tablehead{ \colhead{Name}    & \colhead{$z$}     & \colhead{$\left. \da\right|_{\rm
ellip} ^{\ a}$}     & \colhead{$\left. \da\right|_{\rm sphe} ^{\ b}$}     &
\colhead{$\left. \da\right|_{\rm sphe}^{\textrm{cons}\ c}$}       & \colhead{$\left.
\da\right|_{\rm sphe}^{\textrm{corr}\ d}$}       & \colhead{num$|_{\rm ovl} ^{\ e_1}$} &
\colhead{num$|_{\rm ellip} ^{\ e_2}$}      & \colhead{num$|_{\rm sphe} ^{\ e_3}$}
\\
& & \colhead{Mpc} & \colhead{Mpc} & \colhead{Mpc} & \colhead{Mpc} & & & } \startdata
\multicolumn{9}{c}{Overlap Sample}\tabularnewline \hline\noalign{\smallskip}

MS~1137.5+6625  & 0.784  & $2479\pm1023$   & $5070^{+1960}_{-1430}$& $5070\pm1960$ & $5467.5\pm1695$   & 2  & 2    & 2   \\
MS~0451.6-0305  & 0.550  & $1073\pm238$    & $1470^{+270}_{-230}$  & $1470\pm270 $ & $1500\pm250 $   &   5  & 5    & 5   \\
CL~0016+1609    & 0.541$^f$  & $1635\pm391$& $1220^{+220}_{-190}$  & $1220\pm220 $ & $1242.5\pm205 $   & 4  & 4    & 3     \\
RX~J1347.5-1145 & 0.451  & $1166\pm262$    & $510^{+120}_{-110}$   & $ 510\pm120 $ & $ 517.5\pm115 $   & 7  & 7    & 7       \\
Abell~370       & 0.374$^f$  & $1231\pm441$& $1830^{+410}_{-380}$  & $1830\pm410 $ & $1852.5\pm395 $   & 2  & 2    & 2   \\
MS 1358.4+6245  & 0.327  & $697\pm183$     & $810^{+280}_{-230}$   & $ 810\pm280 $ & $ 847.5\pm255 $   & 3  & 3    & 3     \\
Abell~1995      & 0.322  & $885\pm207$     & $1200^{+210}_{-160}$  & $1200\pm210 $ & $1237.5\pm185 $   & 3  & 3    & 3    \\
Abell~611       & 0.288  & $934\pm331$     & $830^{+220}_{-190}$   & $ 830\pm220 $ & $ 852.5\pm205 $   & 5  & 5    & 3     \\
Abell~697       & 0.282  & $1099\pm308$    & $770^{+210}_{-170}$   & $ 770\pm210 $ & $ 800\pm190 $   &   6  & 6    & 6   \\
Abell~1835      & 0.252  & $946\pm131$     & $690^{+160}_{-90}$    & $ 690\pm160 $ & $ 742.5\pm125 $   & 6  & 6    & 5    \\
Abell~2261      & 0.224  & $1118\pm283$    & $950^{+300}_{-260}$   & $ 950\pm300 $ & $ 980\pm280 $   &   1  & 1    & 1   \\
Abell~773       & 0.217$^f$  & $1465\pm407$& $1560^{+360}_{-350}$  & $1560\pm360 $ & $1567.5\pm355 $   & 12 & 12   & 12      \\
Abell~2163      & 0.202  & $806\pm163$     & $730^{+270}_{-220}$   & $ 730\pm270 $ & $ 767.5\pm245 $   & 5  & 5$^{h_1}$& 5      \\
Abell~1689      & 0.183  & $604\pm84$      & $900^{+160}_{-190}$   & $ 900\pm190 $ & $ 877.5\pm175 $   & 5  & 5    & 5      \\
Abell~665       & 0.182  & $451\pm189$     & $760^{+160}_{-150}$   & $ 760\pm160 $ & $ 767.5\pm155 $   & 5  & 5    & 5      \\
Abell~2218      & 0.176$^f$  & $809\pm263$ & $1180^{+240}_{-220}$  & $1180\pm240 $ & $1195\pm230 $   &   5  & 5    & 3   \\
Abell~1413      & 0.142  & $478\pm126$     & $620^{+190}_{-160}$   & $ 620\pm190 $ & $ 642.5\pm175 $   & 5  & 5    & 5      \\

\hline\noalign{\smallskip} \multicolumn{9}{c}{Elliptical $\beta$-model
only}\tabularnewline \hline\noalign{\smallskip}

Abell~520            & 0.202  & $387\pm141$      &   &   &   &   & 5$^{h_1}$&     \\
Abell~2142           & 0.091  & $335\pm70$       &   &   &   &   & 4    &     \\
Abell~478            & 0.088  & $448\pm185$      &   &   &   &   & 4    &    \\
Abell~1651           & 0.084  & $749\pm385$      &   &   &   &   & 1    &    \\
Abell~401            & 0.074  & $369\pm62$       &   &   &   &   & 4    &    \\
Abell~399            & 0.072  & $165\pm45$       &   &   &   &   & 5    &      \\
Abell~2256           & 0.058  & $242\pm61$       &   &   &   &   & 9    &     \\
Abell~1656           & 0.023  & $103\pm42$       &   &   &   &   & 50   &     \\

\hline\noalign{\smallskip} \multicolumn{9}{c}{Spherical $\beta$-model
only}\tabularnewline \hline\noalign{\smallskip}

CL~J1226.9+3332& 0.890      &     & $810^{+280}_{-220}$ & $810\pm280$ &   $855\pm250$ &     &   &   0$^g$   \\
MS~1054.5-0321& 0.826       &     & $1580^{+420}_{-320}$ & $1580\pm420$ & $1655\pm370$ &    &   &   3    \\
RX~J1716.4+6708& 0.813      &     & $2090^{+1070}_{-800}$ & $2090\pm1070$ &$2292.5\pm935$&  &   &   6       \\
MACS~J0744.8+3927 & 0.686   &     & $1830^{+430}_{-410}$ & $1830\pm430$ & $1845\pm420$ &    &   &   4        \\
MACS~J0647.7+7015 & 0.584   &     & $730^{+200}_{-170}$ &  $730\pm200$ &  $752.5\pm185$ &   &   &   8$^{h_2}$   \\
MS~2053.7-0449& 0.583       &     & $3580^{+1620}_{-1240}$ &$3580\pm1620$ &$3865\pm1430$ &  &   &   8$^{h_2}$   \\
MACS~J2129.4-0741& 0.570    &     & $1220^{+340}_{-280}$ & $1220\pm340$ & $1265\pm310$ &    &   &   4     \\
MACS~J1423.8+2404& 0.545    &     & $1710^{+650}_{-570}$ & $1710\pm650$ & $1770\pm610$ &    &   &   1$^{h_3}$  \\
MACS~J1149.5+2223 & 0.544   &     & $1560^{+400}_{-320}$ & $1560\pm400$ & $1620\pm360$ &    &   &   1$^{h_3}$  \\
MACS~J1311.0-0310 & 0.490   &     & $1500^{+760}_{-500}$ & $1500\pm760$ & $1695\pm630$ &    &   &   2     \\
MACS~J2214.9-1359& 0.483    &     & $1860^{+420}_{-340}$ & $1860\pm420$ & $1920\pm380$ &    &   &   2   \\
MACS~J2228.5+2036& 0.412    &     & $1990^{+470}_{-440}$ & $1990\pm470$ & $2012.5\pm455$ &  &   &   6      \\
ZW~3146  & 0.291            &     & $760^{+190}_{-180}$ & $760\pm190$ & $767.5\pm185$ &     &   &   3      \\
Abell~68 & 0.255            &     & $680^{+270}_{-220}$ & $680\pm270$ & $717.5\pm245$ &     &   &   6      \\
RX~J2129.7+0005  & 0.235    &     & $560^{+210}_{-160}$ & $560\pm210$ & $597.5\pm185$ &     &   &   2      \\
Abell~267  & 0.230          &     & $1140^{+370}_{-280}$ & $1140\pm370$ & $1207.5\pm325$ &  &   &   1$^{h_4}$  \\
Abell~2111 & 0.229          &     & $720^{+350}_{-280}$ & $720\pm350$ & $772.5\pm315$ &     &   &   1$^{h_4}$  \\
Abell~586  & 0.171          &     & $740^{+170}_{-220}$ & $740\pm220$ & $702.5\pm195$ &     &   &   2$^{h_5}$  \\
Abell~1914 & 0.171          &     & $670^{+120}_{-130}$ & $670\pm130$ & $662.5\pm125$ &     &   &   2$^{h_5}$  \\
Abell~2259 & 0.164          &     & $510^{+370}_{-290}$ & $510\pm370$ & $570\pm330$ &       &   &   3      \\
Abell~2204 & 0.152          &     & $460^{+110}_{-100}$ & $460\pm110$ & $467.5\pm105$ &     &   &   5      \\

\enddata
\tablecomments{The overlap sample are the galaxy clusters studied by \citet{defi05} using
elliptical $\beta$-model and meanwhile by \citet{bona06} using spherical $\beta$-model.}
\tablenotetext{a}{Computed estimate for the angular diameter distance reported by
\citet{defi05} assuming ellipsoidal geometry for galaxy clusters.}
\tablenotetext{b}{Calculated angular diameter distance reported by \citet{bona06} under
spherical symmetry for galaxy clusters.} \tablenotetext{c}{Conservative angular diameter
distance estimate for spherical $\beta$-model calculated in this work as described in
section 2.} \tablenotetext{d}{Corrected angular diameter distance estimate for spherical
$\beta$-model calculated in this work as described in section 2.}
\tablenotetext{e}{Number of SNe Ia selected for each galaxy cluster in method (II) as
described in section 2. num$|_{\rm ovl}$, num$|_{\rm ellip}$ and num$|_{\rm sphe}$ stand
for the counts of selected SNe Ia for the overlap sample, the whole elliptical sample and
the whole spherical sample, respectively.} \tablenotetext{f}{The two samples by
\citet{defi05} and \citet{bona06} give different redshift values for the same galaxy
clusters, and therefore the final adopted redshift results are taken from the original
observations. The redshift of CL~0016+1609 is from \citet{1991ApJS...76..813S} and those
of Abell~370, 773 \& 2218 from \citet{1999ApJS..125...35S}.} \tablenotetext{g}{There is
no SN Ia satisfying the criterion, $\Delta z=| z_{\rm cluster} - z_{\rm SN}| <0.005$, for
galaxy cluster CL~J1226.9+3332, and thus it is removed from our analysis.}
\tablenotetext{h}{In order to avoid double counting of SNe Ia data as well as consider
all possible clusters data in our analysis, each of the five pairs of clusters, marked by
$h_1,~h_2,~h_3,~h_4,\textrm{and}~h_5$ respectively, has to be binned within, and the
members of each pair have to share the same set of SNe Ia in method (II) as shown. The
binning algorithm is inverse variance weighted binning, similar to Eq.~\ref{eq:dlsigdl}.
For a complete comparison, they are also binned in method (I) analysis.}

\end{deluxetable}

\begin{deluxetable}{lrrrcrc}\label{table:eta}
\tablecolumns{7}
\tabletypesize{\scriptsize} \tablewidth{0pt}

\tablecaption{Summary of one-dimensional maximum likelihood estimation results: $\eta_0$.
The $\eta_0$ below is represented by the best-fit value at 1-$\sigma$ Confidence Level
for each model using two methods.}
\tablehead{ \colhead{} & \multicolumn{2}{c}{Elliptical $\beta$-model} &
\multicolumn{2}{c}{Conservative Spherical $\beta$-model} & \multicolumn{2}{c}{Corrected
Spherical $\beta$-model}
\\
& \multicolumn{2}{c}{\hrulefill} & \multicolumn{2}{c}{\hrulefill} &
\multicolumn{2}{c}{\hrulefill}
\\
\colhead{Parameterization} & \colhead{Method (I)} & \colhead{Method (II)} &
\colhead{Method (I)} & \colhead{Method (II)} & \colhead{Method (I)} & \colhead{Method
(II)} } \startdata \multicolumn{7}{c}{Whole Sample}\tabularnewline
\hline\noalign{\smallskip}

$\eta=1+\eta_0z$           & $-0.063^{+0.175}_{-0.175}$ & $-0.047^{+0.178}_{-0.178}$ & $-0.227^{+0.102}_{-0.102}$ & $-0.266^{+0.102}_{-0.102}$ & $-0.156^{+0.094}_{-0.094}$ & $-0.201^{+0.094}_{-0.094}$ \\
$\eta=1+\eta_0z/(1+z)$     & $-0.103^{+0.242}_{-0.242}$ & $-0.083^{+0.246}_{-0.246}$ & $-0.339^{+0.152}_{-0.152}$ & $-0.396^{+0.153}_{-0.153}$ & $-0.231^{+0.141}_{-0.141}$ & $-0.297^{+0.142}_{-0.142}$ \\
$\eta=1+\eta_0z/(1+z)^2$   & $-0.157^{+0.326}_{-0.326}$ & $-0.132^{+0.331}_{-0.331}$ & $-0.476^{+0.223}_{-0.223}$ & $-0.557^{+0.225}_{-0.225}$ & $-0.319^{+0.206}_{-0.206}$ & $-0.412^{+0.209}_{-0.209}$ \\
$\eta=1-\eta_0\rm{ln}(1+z)$& $ 0.082^{+0.208}_{-0.208}$ & $ 0.064^{+0.211}_{-0.211}$ & $ 0.282^{+0.126}_{-0.126}$ & $ 0.329^{+0.126}_{-0.126}$ & $ 0.193^{+0.116}_{-0.116}$ & $ 0.248^{+0.117}_{-0.117}$ \\

\hline\noalign{\smallskip} \multicolumn{7}{c}{Overlap Sample}\tabularnewline
\hline\noalign{\smallskip}

$\eta=1+\eta_0z$           & $0.021 ^{+0.179}_{-0.179}$ & $0.031 ^{+0.182}_{-0.182}$ & $-0.362^{+0.145}_{-0.145}$ & $-0.378^{+0.147}_{-0.147}$ & $-0.287^{+0.136}_{-0.136}$ & $-0.306^{+0.138}_{-0.138}$ \\
$\eta=1+\eta_0z/(1+z)$     & $0.037 ^{+0.251}_{-0.251}$ & $0.045 ^{+0.255}_{-0.255}$ & $-0.497^{+0.209}_{-0.209}$ & $-0.522^{+0.211}_{-0.211}$ & $-0.387^{+0.194}_{-0.194}$ & $-0.417^{+0.197}_{-0.197}$ \\
$\eta=1+\eta_0z/(1+z)^2$   & $0.064 ^{+0.343}_{-0.343}$ & $0.067 ^{+0.348}_{-0.348}$ & $-0.654^{+0.297}_{-0.297}$ & $-0.691^{+0.299}_{-0.299}$ & $-0.496^{+0.275}_{-0.275}$ & $-0.541^{+0.278}_{-0.278}$ \\
$\eta=1-\eta_0\rm{ln}(1+z)$& $-0.028^{+0.214}_{-0.214}$ & $-0.037^{+0.217}_{-0.217}$ & $0.428 ^{+0.176}_{-0.176}$ & $ 0.449^{+0.178}_{-0.178}$ & $ 0.337^{+0.163}_{-0.163}$ & $0.361 ^{+0.166}_{-0.166}$ \\

\enddata
\end{deluxetable}

\begin{deluxetable}{lccrcrc}
\tablecolumns{7} \tabletypesize{\scriptsize} \tablewidth{0pt}

\tablecaption{Summary of one- and two-dimensional maximum likelihood estimation results:
minimum of reduced chi square ($\chi^2_{\rm min}/{\rm d. o. f.}$). The values below
represent the $\chi^2_{\rm min}/{\rm d. o. f.}$ for each model using two methods.}
\label{table:chi2}

\tablehead{ \colhead{} & \multicolumn{2}{c}{Elliptical $\beta$-model} &
\multicolumn{2}{c}{Conservative Spherical $\beta$-model} & \multicolumn{2}{c}{Corrected
Spherical $\beta$-model}
\\
& \multicolumn{2}{c}{\hrulefill} & \multicolumn{2}{c}{\hrulefill} &
\multicolumn{2}{c}{\hrulefill}
\\
\colhead{Parameterization} & \colhead{Method (I)} & \colhead{Method (II)} &
\colhead{Method (I)} & \colhead{Method (II)} & \colhead{Method (I)} & \colhead{Method
(II)} }

\startdata

\multicolumn{7}{c}{Whole Sample}\tabularnewline \hline\noalign{\smallskip}

$\eta=1+\eta_0z$           & 24.745/23 & 22.031/23 & 68.916/32 & 66.265/32 & 81.575/32 & 77.514/32 \\
$\eta=1+\eta_0z/(1+z)$     & 24.696/23 & 21.989/23 & 68.983/32 & 66.422/32 & 81.640/32 & 77.678/32 \\
$\eta=1+\eta_0z/(1+z)^2$   & 24.645/23 & 21.943/23 & 69.370/32 & 66.971/32 & 81.936/32 & 78.151/32 \\
$\eta=1-\eta_0\rm{ln}(1+z)$& 24.719/23 & 22.009/23 & 68.903/32 & 66.290/32 & 81.573/32 & 77.553/32 \\

$\eta=\eta_1+\eta_2z$           & 24.126/22 & 21.566/22 & 67.809/31 & 64.947/31 & 80.181/31 & 75.829/31 \\
$\eta=\eta_1+\eta_2z/(1+z)$     & 24.104/22 & 21.576/22 & 66.525/31 & 63.528/31 & 78.866/31 & 74.334/31 \\
$\eta=\eta_1+\eta_2z/(1+z)^2$   & 24.069/22 & 21.588/22 & 64.748/31 & 61.592/31 & 76.988/31 & 72.241/31 \\
$\eta=\eta_1-\eta_2\rm{ln}(1+z)$& 24.118/22 & 21.573/22 & 67.150/31 & 64.219/31 & 79.508/31 & 75.066/31 \\

\hline\noalign{\smallskip} \multicolumn{7}{c}{Overlap Sample}\tabularnewline
\hline\noalign{\smallskip}

$\eta=1+\eta_0z$           & 13.980/16 & 12.725/16 & 48.284/16 & 46.460/16 & 56.947/16 & 54.320/16 \\
$\eta=1+\eta_0z/(1+z)$     & 13.972/16 & 12.722/16 & 48.837/16 & 46.971/16 & 57.464/16 & 54.800/16 \\
$\eta=1+\eta_0z/(1+z)^2$   & 13.960/16 & 12.717/16 & 49.626/16 & 47.735/16 & 58.170/16 & 55.493/16 \\
$\eta=1-\eta_0\rm{ln}(1+z)$& 13.976/16 & 12.724/16 & 48.537/16 & 46.689/16 & 57.186/16 & 54.536/16 \\

$\eta=\eta_1+\eta_2z$           & 13.832/15 & 12.683/15 & 42.376/15 & 41.302/15 & 50.079/15 & 48.419/15 \\
$\eta=\eta_1+\eta_2z/(1+z)$     & 13.811/15 & 12.673/15 & 40.559/15 & 39.517/15 & 48.033/15 & 46.418/15 \\
$\eta=\eta_1+\eta_2z/(1+z)^2$   & 13.803/15 & 12.664/15 & 38.222/15 & 37.195/15 & 45.378/15 & 43.785/15 \\
$\eta=\eta_1-\eta_2\rm{ln}(1+z)$& 13.819/15 & 12.678/15 & 41.433/15 & 40.378/15 & 49.018/15 & 47.384/15 \\

\enddata
\end{deluxetable}

\end{document}